\documentstyle{article}

\catcode`\@=11
\def\relaxnext@{\let\next\relax}
\font\tenmsy=msym10 scaled\magstep1
\font\sevenmsy=msym7 scaled\magstep1
\font\fivemsy=msym5  scaled\magstep1
\newfam\msyfam
\textfont\msyfam=\tenmsy
\scriptfont\msyfam=\sevenmsy
\scriptscriptfont\msyfam=\fivemsy
\font\teneuf=eufm10 scaled\magstep1
\font\seveneuf=eufm7 scaled\magstep1
\font\fiveeuf=eufm5 scaled\magstep1
\newfam\euffam
\textfont\euffam=\teneuf
\scriptfont\euffam=\seveneuf
\scriptscriptfont\euffam=\fiveeuf
\def\frak{\relaxnext@\ifmmode\let\next\frak@\else
 \def\next{\Err@{Use \string\frak\space only in math mode}}\fi\next}
\def\goth{\relaxnext@\ifmmode\let\next\frak@\else
 \def\next{\Err@{Use \string\goth\space only in math mode}}\fi\next}
\def\frak@#1{{\frak@@{#1}}}
\def\frak@@#1{\noaccents@\fam\euffam#1}
\def\Bbb{\relaxnext@\ifmmode\let\next\Bbb@\else
 \def\next{\Err@{Use \string\Bbb\space only in math mode}}\fi\next}
\def\Bbb@#1{{\Bbb@@{#1}}}
\def\Bbb@@#1{\noaccents@\fam\msyfam#1}
\def\accentfam@{7}
\def\noaccents@{\def\accentfam@{0}}
\catcode`\@=\active

\newcommand{\bz}{{\Bbb Z}}

\newcommand{\bc}{{\Bbb C}}

\makeatletter
\@addtoreset{equation}{section}
\makeatother

\newtheorem{thm}{Theorem}[section]
\newtheorem{prop}[thm]{Proposition}
\newtheorem{lem}[thm]{Lemma}
\newtheorem{cor}[thm]{Corollary}

\newtheorem{df}{Definition}[section]

\begin{document}

\begin{flushright}
Univ. Melbourne \\
preprint No.26 \\
hep-th/9408086 \\
July 1994
\end{flushright}

\vspace{36pt}

\begin{center}
\begin{Large}
Virasoro character identities from the Andrews--Bailey construction

\vspace{24pt}

Omar Foda and Yas-Hiro Quano\raisebox{2mm}{$\star$}

\vspace{24pt}
{\it Department of Mathematics, University of Melbourne \\
 \it Parkville, Victoria 3052, Australia}
\end{Large}
\vspace{24pt}

\underline{ABSTRACT}

\end{center}

\vspace{24pt}
We prove $q$-series identities between bosonic and fermionic
representations of certain Virasoro characters.  These identities
include some of the conjectures made by the Stony Brook group
as special cases. Our method is a direct application of Andrews'
extensions of Bailey's lemma to recently obtained polynomial
identities.

\vfill
\hrule

\vskip 3mm
\begin{small}

\noindent\raisebox{2mm}{$\star$} Supported by
the Australian Research Council.

\end{small}

\newpage

\section{Introduction}

\subsection{Aim}

In an impressive series of papers that include \cite{SB1},
the Stony Brook group conjectured a large number of $q$-series
identities. For reviews and complete references, see \cite{SB2}.
Let us restrict our attention to those conjectures related to
the following Virasoro characters \cite{SB1}:

\begin{tabular}{cl}
(i) & $\chi^{(p,p+1)}_{r,s}(q)$ of
the unitary minimal model
${\cal M} (p, p+1)$, for any $r$ and $s$; \\
(ii) &
$\chi^{(p, p+2)}_{(p-1)/2, (p+1)/2}(q)$
of the non-unitary minimal model
${\cal M} (p, p+2)$, where $p$ is odd; \\
(iii) &
$\chi^{(p, kp+1)}_{1,k}(q)$
of the non-unitary minimal model
${\cal M} (p, kp+1)$  where $p \geq 3$ and
$k \geq 1$.
\end{tabular}

Here, ${\cal M} (p,p')$ is a Virasoro minimal model specified by
two coprime integers $(p,p')$, and $\chi^{(p,p')}_{r,s}(q)$ is
a conformal character of irreducible highest weight representation
of ${\cal M} (p,p')$, specified by two integers $(r,s)$, where
$1 \leq r \leq p-1$, $1 \leq s \leq p'-1$ \cite{ISZ}.

These identities are of great interest for a number of physical
and mathematical reasons which are beyond the scope of this work.
The first step towards proving them was taken by Melzer \cite{Mel1},
who conjectured a polynomial identity which implies the $q$-series
identities for (i) in the above list, and proved it for $p=3,4$.
In \cite{Ber} Berkovich proved these polynomial identities for
arbitrary $p$ and for $s=1$, and thus proved the $q$-series
identities involving (i) for $s=1$. In \cite{FQ} we presented
a polynomial identity which implies Gordon's generalization
of the Rogers--Ramanujan identities \cite{Andbk}, and the $q$-series
identities for $\chi^{(2,2k+1)}_{1,i}(q)$ \cite{FNO,NRT}.

In this paper, we prove $q$-series identities for

\begin{tabular}{cl}
(II) & $\left\{
\begin{tabular}{l}
$\chi^{(2\kappa +1,k(2\kappa +1)+2)}_{
       \iota ,k\kappa +1}(q)$ and
$\chi^{(2\kappa +1,k(2\kappa +1)+2)}_{
       \iota ,(k+1)\kappa +1}(q)$ \\
$\chi^{(2\kappa +1,k(2\kappa +1)+2\kappa -1)}_{
       \iota ,k\kappa +k-1}(q)$ and
$\chi^{(2\kappa +1,k(2\kappa +1)+2\kappa -1)}_{
       \iota ,(k+1)\kappa +k-1}(q)$
\end{tabular} \right.$ ~~
for $1 \leq \iota \leq \kappa $. \\
(III) & $\left\{
\begin{tabular}{l}
$\chi^{(p,kp+p-1)}_{1,r(k+1)}(q)$ and
$\chi^{(p,kp+p-1)}_{1,r(k+1)+k}(q)$ \\
$\chi^{(p,kp+1)}_{1,kr}(q)$ and
$\chi^{(p,kp+1)}_{1,k(r+1)+1}(q)$
\end{tabular} \right.$
for $p \geq 4$, $1 \leq r \leq p-2$ and $k\geq 1$,
\end{tabular}

\noindent which include (ii) and (iii) as special cases,
respectively. In order to obtain the above fermionic
representations for (II) and (III), we apply Andrews' extensions
of Bailey's lemma \cite{Bai,And,AAB,Bre2} to the polynomial
identities presented in \cite{FQ}, and \cite{Mel1,Ber},
respectively.

\subsection{Plan}

This paper is organized as follows. In the rest of this section,
we formulate the problem and summarize our results. In section 2
we review a number of definitions and propositions concerning
Andrews' extensions of Bailey's lemma \cite{Bai,And,AAB,Bre2}.
We wish to refer to these extensions as the {\it Andrews--Bailey
construction}. In section 3 and section 4 we obtain the desired
$q$-series identities related to (II) and (III), respectively.
In section 5 we give some remarks. Appendix A contains a summary
of some of the $q$-series identities that can be obtained by a
direct application of the Andrews--Bailey construction to Slater's
list of Bailey pairs \cite{Sla}.

\subsection{Formulation of the problem}

Rocha-Caridi \cite{RC} obtained the following expression
for the Virasoro characters as an infinite series in $q$:
\begin{equation}
\chi^{(p,p')}_{r,s} (q)= \frac{1}{(q)_{\infty}}
\sum_{n=-\infty}^{\infty}
\left( q^{pp'n^2 +(rp'-sp)n} - q^{(pn+r)(p'n+s)} \right),
\label{eqn:RC}
\end{equation}
where
$p$ and $p'$ are coprime positive integers,
$1 \leq r \leq p-1$, $1 \leq s \leq p'$, and
$$
(a)_{\infty} \equiv (a; q)_{\infty}
=\prod_{m=0}^{\infty}(1-aq^{m}).
$$
Starting from Bethe ansatz computations, the Stony Brook
group \cite{SB1} found different $q$-series expressions for
large classes of these characters. For physical reasons
their new expressions are referred to as the {\it fermionic}
sum representations, whereas (\ref{eqn:RC}) is referred to as
the {\it bosonic} sum representation.

For non-unitary minimal model ${\cal M}(2, 2k+1)$, (\ref{eqn:RC})
reduces to
\begin{equation}
\begin{array}{rcl}
\chi^{(2,2k+1)}_{1,i} (q) & = &
\displaystyle\frac{1}{(q)_{\infty}}
\sum_{n=-\infty}^{\infty}
\left( q^{(4k+2)n^2 +(2k-2i+1)n} - q^{(2n+1)((2k+1)n+i)} \right) \\
& = &
\displaystyle\frac{1}{(q)_{\infty}}
\sum_{n=-\infty}^{\infty}
(-1)^n q^{n((2k+1)n +2k-2i+1)/2} \\
& = &
\displaystyle\prod_{n=1 \atop
                    n \not\equiv 0, \pm i ({\rm mod }2k+1)}^{\infty}
(1-q^n )^{-1},
\end{array}
\label{eqn:2,2k+1}
\end{equation}
where the last equality is obtained using Jacobi's triple product
formula.

The fermionic expression corresponding to (\ref{eqn:2,2k+1}),
which was obtained in \cite{FNO,NRT} from different approaches,
is as follows:
\begin{equation}
\chi^{(2,2k+1)}_{1,i} (q) =
\chi^{(2,2k+1)}_{1,2k+1-i} (q) =
\displaystyle\sum_{n_1 \geq \cdots \geq n_{k-1} \geq 0}
\frac{q^{n_1^2 +\cdots + n_{k-1}^2 +n_i +\cdots +n_{k-1}}}
     {(q)_{n_1 -n_2 } \cdots (q)_{n_{k-2}-n_{k-1}} (q)_{n_{k-1}}},
\end{equation}
where $1 \leq i \leq k$, and
$$
(a)_n \equiv (a; q)_n =
\frac{(a)_{\infty}}{(aq^{n})_{\infty}}.
$$
Equating these two expressions, we reproduce
Gordon's generalization of the Rogers--Ramanujan identity
\cite{Andbk} for $1\leq i \leq k$
\begin{equation}
\prod_{n=1 \atop n \not\equiv 0, \pm i {\rm (mod }2k+1)}^{\infty}
(1-q^n )^{-1}
=
\displaystyle\sum_{n_1 \geq \cdots \geq n_{k-1} \geq 0}
\frac{q^{n_1^2 +\cdots + n_{k-1}^2 +n_i +\cdots +n_{k-1}}}
     {(q)_{n_1 -n_2 } \cdots (q)_{n_{k-2}-n_{k-1}} (q)_{n_{k-1}}}.
\label{eqn:Gor}
\end{equation}
This is the simplest $q$-series identity between bosonic and fermionic
representations of Virasoro characters.

In \cite{SB1} the Stony Brook group proved identities of the above
type up to a finite power in $q$, using explicit computations, and
conjectured their validity to all powers in $q$. Proving some of
these conjectures is the problem we address in this work
\footnote{ There exists an extensive literature devoted to this
topic but not discussed in this paper \cite{KNS,KR,KRV,Mel2}. }.

\subsection{Summary of results}

In section 3 we shall prove the following $q$-series identities.
The notation will be explained in section 3.1.

\begin{thm} The following identities hold:
\begin{equation}
\begin{array}{cl}
&
\displaystyle\frac{1}{(q)_{\infty}}
\sum_{n=-\infty}^{\infty}
\left( q^{(2\kappa +1)((2\kappa +1)k+2)n^2 +
          (\iota ((2\kappa +1)k+2)-(k\kappa +1)(2\kappa +1))n} \right. \\
- &
\left. q^{((2\kappa +1)n+\iota )((2\kappa +1)k+2)n+)(k\kappa +1)} \right) \\
= &
\displaystyle\sum_{n_1 \geq \cdots \geq n_{k}\geq 0}
\frac{q^{n_1^2 +\cdots + n_{k}^2
        +(\kappa -\iota )(n_1 + \cdots + n_{k})}}
     {(q)_{n_1 - n_2 }\cdots (q)_{n_{k-1}-n_{k}}
      (q)_{2n_{k}+\kappa -\iota }} \\
\times &
\displaystyle
\sum_{\nu _1 \geq \cdots \geq \nu _{\kappa -1} \geq \nu _{\kappa }=0 \atop
      \nu _{1} + \cdots + \nu _{\kappa -1} \leq n_{k}}
q^{\nu _1^2 + \cdots + \nu _{\kappa -1}^2
  +\nu _{\iota } + \cdots + \nu _{\kappa -1}} \\
\times &
\displaystyle\prod_{\mu =1}^{\kappa -1}
\left[ \begin{array}{c}
2n_{k} +\kappa -\iota -2(\nu _{1} +
\cdots + \nu _{\mu -1})-\nu _{\mu } -\nu _{\mu +1}
-\alpha^{(\kappa )}_{\iota \mu } \\
\nu _{\mu } -\nu _{\mu +1}
\end{array} \right]_q .
\end{array}
\end{equation}

{}~

\begin{equation}
\begin{array}{cl}
&
\displaystyle\frac{1}{(q)_{\infty}}
\sum_{n=-\infty}^{\infty}
\left( q^{(2\kappa +1)((2\kappa +1)k+2)n^2 +
          (\iota ((2\kappa +1)k+2)-(k(\kappa +1)+1)(2\kappa +1))n} \right. \\
- & \left.
q^{((2\kappa +1)n+\iota )((2\kappa +1)k+2)n+)(k(\kappa +1)+1)} \right) \\
= &
\displaystyle\sum_{n_1 \geq \cdots \geq n_{k}\geq 0}
\frac{q^{n_1^2 \cdots + n_{k}^2 +
(\kappa -\iota +1)(n_1 + \cdots + n_k )}}
 {(q)_{n_1 - n_2 }\cdots (q)_{n_{k-1}-n_{k}}
  (q)_{2n_{k}+\kappa -\iota +1}} \\
\times &
\displaystyle\sum_{
\nu_1 \geq \cdots \geq \nu_{\kappa -1} \geq \nu_{\kappa }=0 \atop
      \nu_{1} + \cdots + \nu_{\kappa -1} \leq n_{k}}
q^{\nu_1^2 + \cdots + \nu_{\kappa -1}^2
+\nu_{\iota } + \cdots + \nu_{\kappa -1}} \\
\times &
\displaystyle
\prod_{\mu =1}^{\kappa -1}
\left[ \begin{array}{c}
2n_{k}+\kappa -\iota +1
-2(\nu_{1} + \cdots + \nu_{\mu -1})-\nu_{\mu } -\nu_{\mu +1}
-\alpha^{(\kappa )}_{\iota \mu } \\
\nu_{\mu } -\nu_{\mu +1}
\end{array} \right]_q .
\end{array}
\end{equation}
\end{thm}

As a corollary we have

\begin{cor}
\begin{equation}
\begin{array}{rcl}
\chi^{(2\kappa +1, (2\kappa +1)k+2)}_{\iota ,k\kappa +1}(q)
& = &
\displaystyle\sum_{n_1 \geq \cdots \geq n_{k}\geq 0}
\frac{q^{n_1^2 +\cdots + n_{k}^2
        +(\kappa -\iota )(n_1 + \cdots + n_{k})}}
     {(q)_{n_1 - n_2 }\cdots (q)_{n_{k-1}-n_{k}}
      (q)_{2n_{k}+\kappa -\iota }} \\
& \times &
\displaystyle
\sum_{\nu _1 \geq \cdots \geq \nu _{\kappa -1} \geq \nu _{\kappa }=0 \atop
      \nu _{1} + \cdots + \nu _{\kappa -1} \leq n_{k}}
q^{\nu _1^2 + \cdots + \nu _{\kappa -1}^2
  +\nu _{\iota } + \cdots + \nu _{\kappa -1}} \\
& \times &
\displaystyle\prod_{\mu =1}^{\kappa -1}
\left[ \begin{array}{c}
2n_{k} +\kappa -\iota -2(\nu _{1} +
\cdots + \nu _{\mu -1})-\nu _{\mu } -\nu _{\mu +1}
-\alpha^{(\kappa )}_{\iota \mu } \\
\nu _{\mu } -\nu _{\mu +1}
\end{array} \right]_q .
\end{array}
\label{eqn:k,kp+2,1}
\end{equation}

\begin{equation}
\begin{array}{rcl}
\chi^{(2\kappa +1, (2\kappa +1)k+2)}_{\iota ,k(\kappa +1)+1}(q)
& = &
\displaystyle\sum_{n_1 \geq \cdots \geq n_{k}\geq 0}
\frac{q^{n_1^2 \cdots + n_{k}^2 +
(\kappa -\iota +1)(n_1 + \cdots + n_k )}}
 {(q)_{n_1 - n_2 }\cdots (q)_{n_{k-1}-n_{k}}
  (q)_{2n_{k}+\kappa -\iota +1}} \\
& \times &
\displaystyle\sum_{
\nu_1 \geq \cdots \geq \nu_{\kappa -1} \geq \nu_{\kappa }=0 \atop
      \nu_{1} + \cdots + \nu_{\kappa -1} \leq n_{k}}
q^{\nu_1^2 + \cdots + \nu_{\kappa -1}^2
+\nu_{\iota } + \cdots + \nu_{\kappa -1}} \\
\times &
\displaystyle
\prod_{\mu =1}^{\kappa -1} &
\displaystyle\left[ \begin{array}{c}
2n_{k}+\kappa -\iota +1
-2(\nu_{1} + \cdots + \nu_{\mu -1})-\nu_{\mu } -\nu_{\mu +1}
-\alpha^{(\kappa )}_{\iota \mu } \\
\nu_{\mu } -\nu_{\mu +1}
\end{array} \right]_q .
\end{array}
\end{equation}
\end{cor}

Setting $k=1, \iota =\kappa =(p-1)/2$ and
$2n_1 =m_1 , 2\nu_1 =m_1 -m_2 , \cdots , 2\nu_{\kappa -1} =
m_{\kappa -1}-m_{\kappa }$
in (\ref{eqn:k,kp+2,1}), we reproduce the corresponding expressions
in \cite{SB1}.

\begin{thm} The following identities hold:
\begin{equation}
\begin{array}{cl}
&
\displaystyle\frac{1}{(q)_{\infty}}
\sum_{n=-\infty}^{\infty}
\left( q^{(2\kappa +1)((2\kappa +1)k+2\kappa -1)n^2 +
          (\iota ((2\kappa +1)k+2\kappa -1)-(k\kappa +k-1)(2\kappa +1))n}
\right. \\
- & \left.
q^{((2\kappa +1)n+\iota )((2\kappa +1)k+2\kappa -1)n+)(k\kappa +k-1)}
\right) \\
= &
\displaystyle\sum_{n_1 \geq \cdots \geq n_{k}\geq 0}
\frac{q^{n_1^2 +\cdots + n_{k-1}^2 +2n_{k}^2
        +(\kappa -\iota )(n_1 + \cdots + n_{k-1} +2n_{k})}}
     {(q)_{n_1 - n_2 }\cdots (q)_{n_{k-1}-n_{k}}
      (q)_{2n_{k}+\kappa -\iota }} \\
\times &
\displaystyle\sum_{
\nu _1 \geq \cdots \geq \nu _{\kappa -1} \geq \nu _{\kappa }=0 \atop
\nu _{1} + \cdots + \nu _{\kappa -1} \leq n_{k}}
q^{\nu _1^2 + \cdots + \nu _{\kappa -1}^2 -
   \nu _1 (2n_k + \kappa -\iota )} \times \\
\times & \displaystyle\prod_{\mu =1}^{\kappa -1}
\displaystyle\left[ \begin{array}{c}
2n_{k} +\kappa -\iota -2(\nu_{1} +
\cdots + \nu _{\mu -1})-\nu _{\mu } -\nu _{\mu +1}
-\alpha^{(\kappa )}_{\iota \mu } \\
\nu _{\mu } -\nu _{\mu +1}
\end{array} \right]_q .
\end{array}
\end{equation}

{}~

\begin{equation}
\begin{array}{cl}
&
\displaystyle\frac{1}{(q)_{\infty}}
\sum_{n=-\infty}^{\infty}
\left( q^{(2\kappa +1)((2\kappa +1)k+2\kappa -1)n^2 +
          (\iota ((2\kappa +1)k+2\kappa -1)-(k\kappa +k+\kappa )
          (2\kappa +1))n} \right. \\
- & \left.
q^{((2\kappa +1)n+\iota )
   ((2\kappa +1)k+2\kappa -1)n+k\kappa +k+\kappa )}
\right) \\
= &
\displaystyle\sum_{n_1 \geq \cdots \geq n_{k}\geq 0}
\frac{q^{n_1^2 \cdots + n_{k-1}^2 + 2n_k^2 +
(\kappa -\iota +1)(n_1 + \cdots + n_{k-1} +2n_k )}}
 {(q)_{n_1 - n_2 }\cdots (q)_{n_{k-1}-n_{k}}
  (q)_{2n_{k}+\kappa -\iota +1}} \\
\times &
\displaystyle\sum_{
\nu_1 \geq \cdots \geq \nu_{\kappa -1} \geq \nu_{\kappa }=0 \atop
      \nu_{1} + \cdots + \nu_{\kappa -1} \leq n_{k}}
q^{\nu_1^2 + \cdots + \nu_{\kappa -1}^2
-\nu_1 (\kappa -\iota +1)} \\
\times &
\displaystyle
\prod_{\mu =1}^{\kappa -1}
\left[ \begin{array}{c}
2n_{k}+\kappa -\iota +1
-2(\nu_{1} + \cdots + \nu_{\mu -1})-\nu_{\mu } -\nu_{\mu +1}
-\alpha^{(\kappa )}_{\iota \mu } \\
\nu_{\mu } -\nu_{\mu +1}
\end{array} \right]_q .
\end{array}
\end{equation}
\end{thm}

As a corollary we have

\begin{cor}
\begin{equation}
\begin{array}{rcl}
\chi^{(2\kappa +1, (2\kappa +1)(k+1)-2)}_{
      \iota, \kappa k +\kappa -1}(q)
& = &
\displaystyle\sum_{n_1 \geq \cdots \geq n_{k}\geq 0}
\frac{q^{n_1^2 +\cdots + n_{k-1}^2 +2n_{k}^2
        +(\kappa -\iota )(n_1 + \cdots + n_{k-1} +2n_{k})}}
     {(q)_{n_1 - n_2 }\cdots (q)_{n_{k-1}-n_{k}}
      (q)_{2n_{k}+\kappa -\iota }} \\
& \times &
\displaystyle\sum_{
\nu _1 \geq \cdots \geq \nu _{\kappa -1} \geq \nu _{\kappa }=0 \atop
\nu _{1} + \cdots + \nu _{\kappa -1} \leq n_{k}}
q^{\nu _1^2 + \cdots + \nu _{\kappa -1}^2 -
   \nu _1 (2n_k + \kappa -\iota )} \times \\
\times & \displaystyle\prod_{\mu =1}^{\kappa -1} &
\displaystyle\left[ \begin{array}{c}
2n_{k} +\kappa -\iota -2(\nu_{1} +
\cdots + \nu _{\mu -1})-\nu _{\mu } -\nu _{\mu +1}
-\alpha^{(\kappa )}_{\iota \mu } \\
\nu _{\mu } -\nu _{\mu +1}
\end{array} \right]_q .
\end{array}
\end{equation}

\begin{equation}
\begin{array}{rcl}
\chi^{(2\kappa +1, (2\kappa +1)k+2\kappa -1)}_{
       \iota ,k(\kappa +1)+\kappa }(q)
& = &
\displaystyle\sum_{n_1 \geq \cdots \geq n_{k}\geq 0}
\frac{q^{n_1^2 \cdots + n_{k-1}^2 + 2n_k^2 +
(\kappa -\iota +1)(n_1 + \cdots + n_{k-1} +2n_k )}}
 {(q)_{n_1 - n_2 }\cdots (q)_{n_{k-1}-n_{k}}
  (q)_{2n_{k}+\kappa -\iota +1}} \\
& \times &
\displaystyle\sum_{
\nu_1 \geq \cdots \geq \nu_{\kappa -1} \geq \nu_{\kappa }=0 \atop
      \nu_{1} + \cdots + \nu_{\kappa -1} \leq n_{k}}
q^{\nu_1^2 + \cdots + \nu_{\kappa -1}^2
-\nu_1 (\kappa -\iota +1)} \\
\times &
\displaystyle
\prod_{\mu =1}^{\kappa -1} &
\displaystyle\left[ \begin{array}{c}
2n_{k}+\kappa -\iota +1
-2(\nu_{1} + \cdots + \nu_{\mu -1})-\nu_{\mu } -\nu_{\mu +1}
-\alpha^{(\kappa )}_{\iota \mu } \\
\nu_{\mu } -\nu_{\mu +1}
\end{array} \right]_q .
\end{array}
\end{equation}
\end{cor}

In section 4 we prove the following $q$-series identities. The
notation will be explained in section 4.1.

\begin{thm}
\begin{equation}
\begin{array}{cl}
&
\displaystyle\frac{1}{(q)_{\infty}}
\sum_{n=-\infty}^{\infty}
\left( q^{p(kp+p-1)n^2 +(kp+p-1-r(k+1)p)n}
- q^{(pn+1)((kp+p-1)n+r(k+1))} \right) \\
= &
\displaystyle\sum_{n_1 \geq \cdots \geq n_{k}\geq 0}
\frac{q^{n_1^2 +\cdots + n_{k}^2
        +(r-1)(n_1 + \cdots + n_{k})}}
     {(q)_{n_1 - n_2 }\cdots (q)_{n_{k-1}-n_{k}}
      (q)_{2n_{k}+r-1}} \\
\times &
\displaystyle\sum_{{\bf m}\in 2 \bz^n +{\bf Q}_{p-1-r,p-1}}
q^{({\bf m}, C_{p-3} {\bf m})/4 }
\prod_{a=1}^{p-3}
\left[
\begin{array}{c}
({\bf e}_a , I_{p-3} {\bf m} + {\bf e}_{p-1-r} +(2n_k +r-1){\bf e}_1)/2 \\
({\bf e}_a , {\bf m})
\end{array}
\right]_q .
\end{array}
\end{equation}

\begin{equation}
\begin{array}{cl}
&
\displaystyle\frac{1}{(q)_{\infty}}
\sum_{n=-\infty}^{\infty}
\left( q^{p(kp+p-1)n^2 +(kp+p-1-(rk+k+r)p)n}
-
q^{(pn+1)((kp+p-1)n+rk+k+r)} \right) \\
= &
\displaystyle\sum_{n_1 \geq \cdots \geq n_{k}\geq 0}
\frac{q^{n_1^2 +\cdots + n_{k}^2
        +r(n_1 + \cdots + n_{k})}}
     {(q)_{n_1 - n_2 }\cdots (q)_{n_{k-1}-n_{k}}
      (q)_{2n_{k}+r}} \\
\times &
\displaystyle\sum_{{\bf m}\in 2 \bz^n +{\bf Q}_{r,1}}
q^{({\bf m}, C_{p-3} {\bf m})/4 }
\prod_{a=1}^{p-3}
\left[
\begin{array}{c}
({\bf e}_a , I_{p-3} {\bf m} + {\bf e}_{r} +(2n_k +r){\bf e}_1)/2 \\
({\bf e}_a , {\bf m})
\end{array}
\right]_q .
\end{array}
\end{equation}
\end{thm}

As a corollary we have

\begin{cor}
\begin{equation}
\begin{array}{rcl}
\chi^{(p, kp+p-1)}_{1,r(k+1)}(q)
& = &
\displaystyle\sum_{n_1 \geq \cdots \geq n_{k}\geq 0}
\frac{q^{n_1^2 +\cdots + n_{k}^2
        +(r-1)(n_1 + \cdots + n_{k})}}
     {(q)_{n_1 - n_2 }\cdots (q)_{n_{k-1}-n_{k}}
      (q)_{2n_{k}+r-1}} \\
& \times &
\displaystyle\sum_{{\bf m}\in 2 \bz^n +{\bf Q}_{p-1-r,p-1}}
q^{({\bf m}, C_{p-3} {\bf m})/4 } \\
& \times &
\displaystyle\prod_{a=1}^{p-3}
\left[
\begin{array}{c}
({\bf e}_a , I_{p-3} {\bf m} + {\bf e}_{p-1-r} +(2n_k +r-1){\bf e}_1)/2 \\
({\bf e}_a , {\bf m})
\end{array}
\right]_q .
\end{array}
\end{equation}

\begin{equation}
\begin{array}{rcl}
\chi^{(p, kp+p-1)}_{1,r(k+1)+k}(q)
& = &
\displaystyle\sum_{n_1 \geq \cdots \geq n_{k}\geq 0}
\frac{q^{n_1^2 +\cdots + n_{k}^2
        +r(n_1 + \cdots + n_{k})}}
     {(q)_{n_1 - n_2 }\cdots (q)_{n_{k-1}-n_{k}}
      (q)_{2n_{k}+r}} \\
& \times &
\displaystyle\sum_{{\bf m}\in 2 \bz^n +{\bf Q}_{r,1}}
q^{({\bf m}, C_{p-3} {\bf m})/4 }
\prod_{a=1}^{p-3}
\left[
\begin{array}{c}
({\bf e}_a , I_{p-3} {\bf m} + {\bf e}_{r} +(2n_k +r){\bf e}_1)/2 \\
({\bf e}_a , {\bf m})
\end{array}
\right]_q .
\end{array}
\end{equation}
\end{cor}

\begin{thm}
\begin{equation}
\begin{array}{cl}
&
\displaystyle\frac{1}{(q)_{\infty}}
\sum_{n=-\infty}^{\infty}
\left( q^{p(kp+1)n^2 +(kp+1-krp)n}
-
q^{(pn+1)((kp+1)n+kr))} \right) \\
= &
\displaystyle\sum_{n_1 \geq \cdots \geq n_{k}\geq 0}
\frac{q^{n_1^2 +\cdots + n_{k-1}^2 + 2n_k^2 +
        +(r-1)(n_1 + \cdots + n_{k-1}+2n_k )}}
     {(q)_{n_1 - n_2 }\cdots (q)_{n_{k-1}-n_{k}}
      (q)_{2n_{k}+r-1}} \\
\times &
\displaystyle\sum_{{\bf m}\in 2 \bz^n +{\bf Q}_{p-1-r,p-1}}
q^{({\bf m}, C_{p-3} {\bf m})/4
  -({\bf e}_{p-1-r}+(2n_k +r-1){\bf e}_1 , {\bf m})/2} \\
\times &
\displaystyle\prod_{a=1}^{p-3}
\left[
\begin{array}{c}
({\bf e}_a , I_{p-3} {\bf m} + {\bf e}_{p-1-r} +(2n_k +r-1){\bf e}_1)/2 \\
({\bf e}_a , {\bf m})
\end{array}
\right]_q .
\end{array}
\end{equation}

\begin{equation}
\begin{array}{cl}
&
\displaystyle\frac{1}{(q)_{\infty}}
\sum_{n=-\infty}^{\infty}
\left( q^{p(kp+1)n^2 +(kp+1-(kr+k+1)p)n}
-
q^{(pn+1)((kp+1)n+kr+k+1)} \right) \\
= &
\displaystyle\sum_{n_1 \geq \cdots \geq n_{k}\geq 0}
\frac{q^{n_1^2 +\cdots + n_{k-1}^2 +2n_k^2
        +r(n_1 + \cdots + n_{k-1} +2n_k )}}
     {(q)_{n_1 - n_2 }\cdots (q)_{n_{k-1}-n_{k}}
      (q)_{2n_{k}+r}} \\
\times &
\displaystyle\sum_{{\bf m}\in 2 \bz^n +{\bf Q}_{r,1}}
q^{({\bf m}, C_{p-3} {\bf m})/4
  -({\bf e}_{r}+(2n_k +r){\bf e}_1 , {\bf m})/2} \\
\times &
\displaystyle\prod_{a=1}^{p-3}
\left[
\begin{array}{c}
({\bf e}_a , I_{p-3} {\bf m} + {\bf e}_{r} +(2n_k +r){\bf e}_1)/2 \\
({\bf e}_a , {\bf m})
\end{array}
\right]_q .
\end{array}
\end{equation}
\end{thm}

As a corollary we have

\begin{cor}
\begin{equation}
\begin{array}{rcl}
\chi^{(p,kp+1)}_{1,kr}(q)
& = &
\displaystyle\sum_{n_1 \geq \cdots \geq n_{k}\geq 0}
\frac{q^{n_1^2 +\cdots + n_{k-1}^2 +2n_k^2
        +(r-1)(n_1 + \cdots + n_{k-1}+2n_k )}}
     {(q)_{n_1 - n_2 }\cdots (q)_{n_{k-1}-n_{k}}
      (q)_{2n_{k}+r-1}} \\
& \times &
\displaystyle\sum_{{\bf m}\in 2 \bz^n +{\bf Q}_{p-1-r,p-1}}
q^{({\bf m}, C_{p-3} {\bf m})/4
  -({\bf e}_{p-1-r}+(2n_k +r-1){\bf e}_1 , {\bf m})/2} \\
& \times &
\displaystyle\prod_{a=1}^{p-3}
\left[
\begin{array}{c}
({\bf e}_a , I_{p-3} {\bf m} + {\bf e}_{p-1-r} +(2n_k +r-1){\bf e}_1)/2 \\
({\bf e}_a , {\bf m})
\end{array}
\right]_q .
\end{array}
\label{eqn:p,kp+1,1}
\end{equation}

\begin{equation}
\begin{array}{rcl}
\chi^{(p,kp+1)}_{1,k(r+1)+1}(q)
& = &
\displaystyle\sum_{n_1 \geq \cdots \geq n_{k}\geq 0}
\frac{q^{n_1^2 +\cdots + n_{k-1}^2 + 2n_k^2
        +r(n_1 + \cdots + n_{k-1}+2n_k )}}
     {(q)_{n_1 - n_2 }\cdots (q)_{n_{k-1}-n_{k}}
      (q)_{2n_{k}+r}} \\
& \times &
\displaystyle\sum_{{\bf m}\in 2 \bz^n +{\bf Q}_{r,1}}
q^{({\bf m}, C_{p-3} {\bf m})/4
  -({\bf e}_{r}+(2n_k +r){\bf e}_1 , {\bf m})/2} \\
& \times &
\displaystyle\prod_{a=1}^{p-3}
\left[
\begin{array}{c}
({\bf e}_a , I_{p-3} {\bf m} + {\bf e}_{r} +(2n_k +r){\bf e}_1)/2 \\
({\bf e}_a , {\bf m})
\end{array}
\right]_q .
\end{array}
\end{equation}
\end{cor}

Setting $r=1$ and
$n_i -n_{i+1} =\nu_i $ ($i=1, \cdots , k-1$),
$2n_k =\nu_k$, $m_a =\nu_{k+a}$ ($a=1, \cdots , p-3$)
in (\ref{eqn:p,kp+1,1}),
we reproduce the corresponding expressions
given in \cite{SB1}.

\section{The Andrews--Bailey construction}

This section is devoted to a review of a number of definitions
and propositions from the pioneering work of Bailey \cite{Bai}
and its extensions by Andrews \cite{And}.

\subsection{A Bailey pair}

\begin{df} Let $\alpha = \{\alpha_n (a,q)\}_{n\geq 0}$ and
$\beta = \{\beta_n (a,q)\}_{n\geq 0}$
be sequences of functions in $a$ and $q$. They form a Bailey
pair relative to $a$, if they satisfy the relation
\begin{equation}
\beta_n =\sum_{r=0}^{n} \frac{\alpha_r }{(q)_{n-r}(aq)_{n+r}}.
\label{eqn:pair}
\end{equation}
\label{df:Bailey}
\end{df}

Note that (\ref{eqn:pair}) has the form
$\beta = M \alpha $,
where $\alpha = \,\!^t(\alpha_0 , \alpha_1 , \alpha_2 , \cdots ),
\beta =\,\!^t(\beta_0 , \beta_1 , \beta_2 , \cdots )$;
and that $M$ is an invertible matrix of infinite size,
because it is a lower triangular matrix
with non-zero diagonal elements.
Using an identity of $q$-hypergeometric series
\cite[eq.(1.4.3)]{GR},
one can prove the following proposition \cite{And}

\begin{prop}
A pair $(\alpha, \beta )$ is a Bailey pair relative to $a$
if and only if
\begin{equation}
\alpha_n =\frac{1-aq^{2n}}{1-a}
\sum_{r=0}^{n} \frac{(a)_{n+r}(-1)^{n-r}
q^{\left( {n-r} \atop {2} \right) }}
{(q)_{n-r}} \beta_r .
\label{eqn:inv}
\end{equation}
\label{prop:inv}
\end{prop}

Note that the RHS of (\ref{eqn:inv}) has no singularity
at $a=1$.

\subsection{A dual Bailey pair}

Given a Bailey pair, Andrews \cite{And} proposed a method to
construct a new Bailey pair as follows:

\begin{prop}
If $\alpha =\{ \alpha_n (a,q)\}_{n\geq 0}$
and $\beta =\{ \beta_n (a,q)\}_{n\geq 0}$
form a Bailey pair, Then the sequences
$A =\{ A_n (a,q)\}_{n\geq 0}$ and $B =\{ B_n (a,q)\}_{n\geq 0}$
defined by
\begin{equation}
\begin{array}{rcl}
A_n (a,q) & = & a^n q^{n^2 }\alpha_n (a^{-1}, q^{-1}), \\
B_n (a,q) & = & a^{-n} q^{-n^2 -n}\beta_n (a^{-1}, q^{-1}), \\
\end{array}
\label{eqn:dual}
\end{equation}
form another Bailey pair relative to $a$.
\label{prop:dual}
\end{prop}

One can prove Proposition (\ref{prop:dual})
by substituting
$\alpha_n (a^{-1}, q^{-1}), \beta_n (a^{-1}, q^{-1})$
into (\ref{eqn:pair}).
We refer to $(A, B)$ as
the dual of $(\alpha , \beta)$ and
{\it vice versa}.

\subsection{A Bailey chain}

Using Saalsch\"{u}tz theorem
\cite[eq.(1.7.2)]{GR}, one can
prove the celebrated Bailey's lemma \cite{Bai}:

\begin{lem} Let $\alpha$ and $\beta$ be a Bailey pair relative to $a$,
and set
\begin{equation}
\alpha' _n =a^n q^{n^2} \alpha_n , ~~~~
\beta' _n =\sum_{r=0}^{n}
\frac{a^r q^{r^2}}{(q)_{n-r}} \beta_r .
\label{eqn:chain}
\end{equation}
Then $\alpha'$ and $\beta'$ is also a Bailey pair relative to $a$.
\label{lem:chain}
\end{lem}

Andrews \cite{And} proposed a prescription to construct an infinite
sequences of Bailey pairs starting from a given pair using Bailey's
lemma.  Given a Bailey pair $(\alpha^{(0)}, \beta^{(0)})$, one can
obtain another Bailey pair $(\alpha^{(1)}, \beta^{(1)})$ by applying
(\ref{eqn:chain}). Repeating this procedure, one obtains
$(\alpha^{(k)}, \beta^{(k)})$
inductively from
$(\alpha^{(k-1)}, \beta^{(k-1)})$.
The resulting infinite sequence of pairs is called a {\it Bailey chain}.

Substituting $\alpha^{(k)}$ and $\beta^{(k)}$ into the defining relation
(\ref{eqn:pair}) to get
$$
\sum_{r=0}^{n}
\frac{a^{k r} q^{k r^2 }\alpha^{(0)}_r }{(q)_{n-r}(aq)_{n+r}}
=
\sum_{n_1 =0}^{n} \sum_{n_2 =0}^{n_1 }
\cdots \sum_{n_{k}=0}^{n_{k-1}}
\frac{a^{n_1 + \cdots n_k} q^{n_1^2 +\cdots + n_k^2 }
      \beta^{(0)}_{n_k}}
     {(q)_{n-n_1 }(q)_{n_1 -n_2 }\cdots (q)_{n_{k-1}-n_{k}}},
$$
and taking the limit $n\rightarrow \infty$, Andrews \cite{And}
obtained the following identity:

\begin{cor}
Let $(\alpha, \beta )$ be a Bailey pair relative to $a$.
The following holds:
\begin{equation}
\frac{1}{(aq)_{\infty}} \sum_{n=0}^{\infty}
a^{kn} q^{kn^2} \alpha_n =
\sum_{n_1 \geq \cdots \geq n_k \geq 0}
\frac{a^{n_1 + \cdots n_k} q^{n_1^2 +\cdots + n_k^2 } \beta_{n_k}}
     {(q)_{n_1 -n_2 }\cdots (q)_{n_{k-1}-n_{k}}}.
\label{eqn:k-fold}
\end{equation}
\label{cor:And}
\end{cor}

The following remark is the central to this work: for certain
$\alpha_n$, we will be able to rewrite the LHS of (\ref{eqn:k-fold})
in the form of (\ref{eqn:RC}), thus obtaining a $q$-series identity
of the type we wish to prove.

\subsection{A Bailey lattice}

In a Bailey chain, the parameter $a$ remains constant throughout the
chain.  In \cite{AAB,Bre2}, yet another extension of Bailey's lemma
was presented, that allows one to vary $a$. The resulting structure
is called a {\it Bailey lattice}. Using $q$-hypergeometric series
identities \cite[eqs.(1.7.2),(2.2.4) and (2.3.4)]{GR} one can prove
the following Proposition \cite{AAB,Bre2}:

\begin{prop}
Let $\alpha$ and $\beta$ be a Bailey pair relative to $a$
and set
\begin{equation}
\begin{array}{rcl}
\alpha' _n & = &
\left\{ \begin{array}{ll} \alpha_0 , & n=0, \\
        \displaystyle (1-a) a^n q^{n^2 -n} \left\{
        \frac{\alpha_n}{1-aq^{2n}} -
        \frac{aq^{2n-2} \alpha_{n-1}}{1-aq^{2n-2}} \right\}, & n>0,
\end{array} \right. \\
\beta' _n & = &
\displaystyle\sum_{r=0}^{n}
\frac{a^r q^{r^2 -r}}{(q)_{n-r}} \beta_r .
\end{array}
\label{eqn:lattice}
\end{equation}
Then $\alpha'$ and $\beta'$ is also a Bailey pair relative to $aq^{-1}$.
\label{prop:lattice}
\end{prop}

Using the above extension, we can construct a new Bailey pair
$(\alpha^{(k)}(aq^{-1}, q), \beta^{(k)}(aq^{-1}, q))$ from a given
pair $(\alpha^{(0)}(a, q), \beta^{(0)}(a, q))$ as follows. In the
first $k-i-1$ steps, we transform $(\alpha^{(0)}(a, q), \beta^{(0)}(a, q))$
by (\ref{eqn:chain}) as before. At the $(k-i)^{\rm th}$ step, we use
(\ref{eqn:lattice}). After that, we transform
$(\alpha^{(k-i)}(aq^{-1}, q), \beta^{(k-i)}(aq^{-1}, q))$ by
(\ref{eqn:chain}) $i$ times. Substituting
$(\alpha^{(k)}(aq^{-1}, q), \beta^{(k)}(aq^{-1}, q))$
into the defining relation (\ref{eqn:pair}), and taking the
limit $n \rightarrow \infty$, we obtain the following Corollary
\cite{AAB}.

\begin{cor}
Let $(\alpha, \beta )$ be a Bailey pair. Then the following identity
holds:
\begin{equation}
\begin{array}{cl}
&
\displaystyle\sum_{n_1 \geq \cdots \geq n_k \geq 0}
\frac{a^{n_1 + \cdots n_k}
      q^{n_1^2 +\cdots + n_k^2 -n_1 - \cdots -n_i } \beta_{n_k}}
     {(q)_{n_1 -n_2 }\cdots (q)_{n_{k-1}-n_{k}}} \\
= &
\displaystyle\frac{1}{(a)_{\infty}} \left\{ \alpha_0 +
(1-a) \sum_{n=1}^{\infty}
\left[ \frac{a^{kn} q^{kn^2 -in} \alpha_n}{1-aq^{2n}}-
       \frac{a^{k(n-1)} q^{k(n-1)^2 -i(n-1)}(aq^{2(n-1)})^{i+1}
             \alpha_{n-1}}{1-aq^{2n-2}} \right]
\right\}.
\end{array}
\label{eqn:k-i}
\end{equation}
\label{cor:AAB}
\end{cor}

\subsection{Example}

As a starting point, let us choose the following pair of sequences
\cite{AAB}
\begin{equation}
\begin{array}{rcl}
\alpha_n ^{(0)} & = &
\displaystyle\left\{ \begin{array}{ll}
1, & n=0, \\
\displaystyle (-1)^n q^{(n^2 -n)/2}
\displaystyle\frac{1-aq^{2n}}{1-a}\frac{(a)_n}{(q)_n}, & n>0,
\end{array} \right. \\
\beta _n ^{(0)} & = & \delta_{n0}.
\end{array}
\end{equation}
{}From Proposition \ref{prop:inv}
$(\alpha^{(0)}, \beta^{(0)})$ is a Bailey pair. In this example
the dual Bailey pair is the original one itself.

Consider the Bailey chain
$\{ (\alpha^{(k)}, \beta^{(k)}) \}_{k\geq 0}$,
for $a=q^j (j=0, 1)$\footnote{
When $j=0$ (resp. $j=1$) the pair
$(\alpha^{(1)}, \beta^{(1)})$ coincides
B(1) (resp. B(3)) in Slater's table \cite{Sla}. }.
Applying Corollary \ref{cor:And}
to $(\alpha^{(0)}, \beta^{(0)})$,
we obtain
\begin{equation}
\displaystyle\frac{1}{(q)_{\infty}}
\sum_{r=-\infty}^{\infty} (-1)^r q^{r((2k+1)r+1-2jk)/2}
=
\displaystyle\sum_{n_1 \geq \cdots \geq n_{k-1} \geq 0}
\frac{q^{n_1^2 +\cdots + n_{k-1}^2 +j(n_1 +\cdots +n_{k-1})}}
     {(q)_{n_1 -n_2 } \cdots (q)_{n_{k-2}-n_{k-1}} (q)_{n_{k-1}}}.
\label{eqn:ex}
\end{equation}
Setting $k=1$, (\ref{eqn:ex}) reduces to
Euler's pentagonal numbers theorem \cite{LL}
\begin{equation}
(q)_{\infty} = 1 +
\sum_{r=1}^{\infty}
(-1)^r ( q^{r(3r-1)/2} +q^{r(3r+1)/2} );
\end{equation}
and setting $k=2$, (\ref{eqn:ex}) reduces to the Rogers--Ramanujan
identity \cite{Andbk}
\begin{equation}
\displaystyle\frac{1}{(q)_{\infty}}
\sum_{r=-\infty}^{\infty} (-1)^r q^{r(5r+1)/2-2jr)}
=
\displaystyle\sum_{n=0}^{\infty}
\frac{q^{n+jn}}{(q)_{n}}.
\label{eqn:R-R}
\end{equation}

Applying Corollary \ref{cor:AAB} to $(\alpha^{(0)}, \beta^{(0)})$
for $a=q$ and replace $i$ by $i-1$, we obtain Gordon's generalization
of Rogers--Ramanujan identity \cite{Andbk}
\begin{equation}
\displaystyle\frac{1}{(q)_{\infty}}
\sum_{r=-\infty}^{\infty}
(-1)^r q^{r((2k+1)r+2k-2i+1)/2}
=
\displaystyle\sum_{n_1 \geq \cdots \geq n_{k-1} \geq 0}
\frac{q^{n_1^2 +\cdots + n_{k-1}^2 +n_i +\cdots +n_{k-1}}}
     {(q)_{n_1 -n_2 } \cdots (q)_{n_{k-2}-n_{k-1}} (q)_{n_{k-1}}},
\label{eqn:Gordon}
\end{equation}
where $1\leq i \leq k$.
By comparing with (\ref{eqn:Gor})
one can see that the LHS and RHS of (\ref{eqn:Gordon})
are the bosonic
and fermionic representations
of $\chi^{(2,2k+1)}_{1,i}(q)$, respectively.

\section{Fermionic sum representations of the type (II)}

In this section we apply the Andrews-Bailey construction
to the polynomial identity obtained in \cite{FQ}.

\subsection{A polynomial identity}

Our starting point is the following polynomial identity
which implies Gordon's generalization of the Rogers--Ramanujan
identities (\ref{eqn:Gordon}):

\begin{prop}
Let $\nu , \kappa, \iota$ be fixed non negative integers such that
$1 \leq \iota \leq \kappa$. Then the following polynomial identities
hold.
\begin{equation}
\begin{array}{cl}
&
\displaystyle\sum_{\rho=-\infty}^{\infty} (-1)^{\rho }
q^{\rho ((2\kappa +1)\rho +2\kappa -2\iota +1)/2}
\left[ \begin{array}{c} \nu \\
\left[ \frac{\nu -\kappa +\iota-(2\kappa +1)\rho }{2} \right]
\end{array} \right]_q \\
= &
\displaystyle\sum_{
           {\nu _1 \geq \cdots \geq \nu _{\kappa -1}
            \geq \nu _{\kappa} =0}
     \atop {2(\nu _1 + \cdots + \nu _{\kappa -1})
            \leq \nu -\kappa +\iota}}
q^{\nu _1^2 + \cdots + \nu _{\kappa -1}^2 +
   \nu _{\iota } + \cdots +\nu _{\kappa -1} } \times \\
\times &
\displaystyle\prod_{\mu =1}^{\kappa -1}
\left[ \begin{array}{c}
\nu -2(\nu _1 + \cdots + \nu _{\mu -1})-\nu _{\mu }
-\nu _{\mu +1} -\alpha^{(\kappa )}_{\iota \mu } \\
\nu _{\mu }-\nu _{\mu +1}
\end{array} \right]_q .
\end{array}
\label{eqn:truncated}
\end{equation}
Here, $[x]$ is the greatest integer part of $x$,
$$
\left[ \begin{array}{c}
N \\ M \end{array}
\right]_q =
\left\{ \begin{array}{ll}
\displaystyle\frac{(q)_N}{(q)_M (q)_{N-M}}, & 0\leq M \leq N, \\
0, & \mbox{otherwise.}
\end{array} \right.
$$
is a {\it $q$-binomial coefficient},
and
$$
\alpha^{(\kappa )}_{\iota \mu } = \left\{
\begin{array}{ll} 0, & \mbox{if $1 \leq \mu \leq \iota -1$,} \\
           \mu -\iota +1,
          & \mbox{if $\iota \leq \mu \leq \kappa -1$}.
\end{array} \right.
$$
\end{prop}

The proof is given in \cite{FQ}. For $\kappa =1, 2$, this identity
appears in \cite{Andbk}.

\subsection{Fermionic sum representations for
${\cal M}(2\kappa +1, (2\kappa +1)k+2)$}

Set $\nu =2n+\kappa -\iota $ and divide (\ref{eqn:truncated})
by $(q^{\kappa -\iota +1})_{2n}$. Then we have
$$
\begin{array}{cl}
&
\displaystyle\sum_{r=0}^{\infty}
\frac{q^{r((4\kappa +2)r+2\kappa -2\iota +1)}}{
      (q)_{n-(2\kappa +1)r} (q^{\kappa -\iota +1})_{n+(2\kappa +1)r}}
+\sum_{r=0}^{\infty}
\frac{q^{r((4\kappa +2)r-2\kappa +2\iota -1)}}{
      (q)_{n+\kappa -\iota -(2\kappa +1)r}
      (q^{\kappa -\iota +1})_{n-\kappa +\iota +(2\kappa +1)r}} \\
- &
\displaystyle\sum_{r=1}^{\infty}
\frac{q^{(2r-1)((2\kappa +1)r-\iota )}}{
      (q)_{n+\kappa -(2\kappa +1)r}
      (q^{\kappa -\iota +1})_{n-\kappa +(2\kappa +1)r}} -
\sum_{r=0}^{\infty}
\frac{q^{(2r+1)((2\kappa +1)r+\iota )}}{
      (q)_{n-\iota -(2\kappa +1)r}
      (q^{\kappa -\iota +1})_{n+\iota +(2\kappa +1)r}} \\
= &
 \displaystyle\frac{1}{(q^{\kappa -\iota +1})_{2n}}
\sum_{ {\nu _1 \geq \cdots \geq \nu _{\kappa -1}
            \geq \nu _{\kappa} =0}
     \atop {\nu _1 + \cdots + \nu _{\kappa -1}
            \leq n} }
q^{\nu _1^2 + \cdots + \nu _{\kappa -1}^2 + \nu _{\iota } + \cdots
   +\nu _{\kappa -1} } \times \\
\times &
\displaystyle\prod_{\mu =1}^{\kappa -1}
\left[ \begin{array}{c}
2n+\kappa -\iota -2(\nu _1 + \cdots + \nu _{\mu -1})-\nu _{\mu }
-\nu _{\mu +1} -\alpha^{(\kappa )}_{\iota \mu } \\
\nu _{\mu }-\nu _{\mu +1}
\end{array} \right]_q .
\end{array}
$$
{}From the above we can read the following Bailey pair relative to
$q^{\kappa -\iota }$:
\begin{equation}
\alpha_n = \left\{
\begin{array}{ll}
1, & n=0, \\
q^{r((4\kappa +2)r+2\kappa -2\iota +1)},
& n=(2\kappa +1)r ~(r \geq 1), \\
q^{r((4\kappa +2)r-2\kappa +2\iota -1)},
& n=(2\kappa +1)r-\kappa +\iota ~(r \geq 1), \\
-q^{(2r-1)((2\kappa +1)r-\iota )},
& n=(2\kappa +1)r-\kappa ~(r \geq 1), \\
-q^{(2r+1)((2\kappa +1)r+\iota )},
& n=(2\kappa +1)r+\iota ~(r \geq 0), \\
0, & \mbox{otherwise},
\end{array} \right.
\label{eqn:L1}
\end{equation}
\begin{equation}
\begin{array}{rcl}
\beta_n & = &
 \displaystyle\frac{1}{(q^{\kappa -\iota +1})_{2n}} \sum_{
           {\nu _1 \geq \cdots \geq \nu _{\kappa -1} \geq \nu _{\kappa }=0}
      \atop{\nu _1 + \cdots + \nu _{\kappa -1} \leq n}}
q^{\nu _1^2 + \cdots + \nu _{\kappa -1}^2 +\nu _{\iota }+
   \cdots + \nu _{\kappa -1} } \times \\
& \times &
 \displaystyle\prod_{\mu =1}^{\kappa -1}
\left[ \begin{array}{c}
2n+\kappa -\iota -
2(\nu _1 + \cdots + \nu _{\mu -1})-\nu _{\mu }-\nu _{\mu +1}
-\alpha^{(\kappa )}_{\iota \mu } \\
\nu _{\mu }-\nu _{\mu +1}
\end{array} \right]_q .
\end{array}
\label{eqn:R1}
\end{equation}
Here and hereafter terms belonging to the same congruence class modulo
$2 \kappa +1$ should be summed up if exists. For instance, when
$\iota =\kappa$
(\ref{eqn:L1}) should read as follows:
\begin{equation}
\alpha_n = \left\{
\begin{array}{ll}
1, & n=0, \\
q^{r((4\kappa +2)r+1)}+
q^{r((4\kappa +2)r-1)},
& n=(2\kappa +1)r ~(r\geq 1), \\
-q^{(2r-1)((2\kappa +1)r-\kappa )},
& n=(2\kappa +1)r-\kappa ~(r \geq 1), \\
-q^{(2r+1)((2\kappa +1)r+\kappa )},
& n=(2\kappa +1)r+\kappa ~(r \geq 0), \\
0, & \mbox{otherwise},
\end{array} \right.
\end{equation}

Noting that
$$
\begin{array}{cl}
& \displaystyle\sum_{n=0}^{\infty}
(q^{\kappa -\iota })^{kn} q^{kn^2} \alpha_{n} \\
= &
1+
\displaystyle\sum_{r=1}^{\infty}
(q^{\kappa -\iota })^{k(2\kappa +1)r}
q^{k(2\kappa +1)^2 r^2}
q^{r(2(2\kappa +1)r+2\kappa -2\iota +1)} \\
+ &
\displaystyle\sum_{r=1}^{\infty}
(q^{\kappa -\iota })^{k((2\kappa +1)r -\kappa +\iota )}
q^{k((2\kappa +1)r -\kappa +\iota )^2}
q^{r(2(2\kappa +1)r-2\kappa +2\iota +1)} \\
- &
\displaystyle\sum_{r=1}^{\infty}
(q^{\kappa -\iota })^{k((2\kappa +1)r -\kappa )}
q^{k((2\kappa +1)r -\kappa )^2}
q^{(2r-1)((2\kappa +1)r-\iota )} \\
- &
\displaystyle\sum_{r=0}^{\infty}
(q^{\kappa -\iota })^{k((2\kappa +1)r -\iota )}
q^{k((2\kappa +1)r -\iota )^2}
q^{(2r+1)((2\kappa +1)r+\iota )} \\
= &
\displaystyle\sum_{r=-\infty}^{\infty}
\left(
q^{r((2\kappa +1)((2\kappa +1)k+2)r+
    (2\kappa +1)(\kappa -\iota +1)-2\iota )}
-
q^{((2\kappa +1)r+\iota )
   (((2\kappa +1)k+2)r +k\kappa +1)}
\right) \\
= &
(q)_{\infty}
\chi^{(2\kappa +1, (2\kappa +1)k +2)}_{
      \iota , k \kappa +1}(q),
\end{array}
$$
and applying Corollary \ref{cor:And} to the Bailey pair
(\ref{eqn:L1}--\ref{eqn:R1}), we obtain the following
fermionic expression for the Virasoro character:
\begin{equation}
\begin{array}{rcl}
\chi^{(2\kappa +1, (2\kappa +1)k+2)}_{\iota ,k\kappa +1}(q)
& = &
\displaystyle\sum_{n_1 \geq \cdots \geq n_{k}\geq 0}
\frac{q^{n_1^2 +\cdots + n_{k}^2
        +(\kappa -\iota )(n_1 + \cdots + n_{k})}}
     {(q)_{n_1 - n_2 }\cdots (q)_{n_{k-1}-n_{k}}
      (q)_{2n_{k}+\kappa -\iota }} \\
& \times &
\displaystyle
\sum_{\nu _1 \geq \cdots \geq \nu _{\kappa -1} \geq \nu _{\kappa }=0 \atop
      \nu _{1} + \cdots + \nu _{\kappa -1} \leq n_{k}}
q^{\nu _1^2 + \cdots + \nu _{\kappa -1}^2
  +\nu _{\iota } + \cdots + \nu _{\kappa -1}} \\
& \times &
\displaystyle\prod_{\mu =1}^{\kappa -1}
\left[ \begin{array}{c}
2n_{k} +\kappa -\iota -2(\nu _{1} +
\cdots + \nu _{\mu -1})-\nu _{\mu } -\nu _{\mu +1}
-\alpha^{(\kappa )}_{\iota \mu } \\
\nu _{\mu } -\nu _{\mu +1}
\end{array} \right]_q .
\end{array}
\label{eqn:k,kp+2,2}
\end{equation}
Setting $k=1, \iota =\kappa =(p-1)/2$ and
$2n_1 =m_1 , 2\nu_1 =m_1 -m_2 ,
\cdots , 2\nu_{\kappa -1} =m_{\kappa -1}-m_{\kappa }$
in (\ref{eqn:k,kp+2,2}),
we reproduce the corresponding expressions
in \cite{SB1}.

Substituting $\nu =2n+\kappa -\iota +1$ into (\ref{eqn:truncated})
and dividing by $(q^{\kappa -\iota +2 })_{2n}$ we obtain another
Bailey pair relative to $q^{\kappa -\iota +1}$
\begin{equation}
\alpha_n = \left\{
\begin{array}{ll}
1, & n=0, \\
q^{r((4\kappa +2)r+2\kappa -2\iota +1)},
& n=(2\kappa +1)r ~(r \geq 1), \\
q^{r((4\kappa +2)r-2\kappa +2\iota -1)},
& n=(2\kappa +1)r-\kappa +\iota -1 ~(r \geq 1), \\
-q^{(2r+1)((2\kappa +1)r+\iota )},
& n=(2\kappa +1)r +\iota ~(r \geq 0), \\
-q^{(2r-1)((2\kappa +1)r-\iota )},
& n=(2\kappa +1)r-\kappa -1 ~(r \geq 1), \\
0, & \mbox{otherwise},
\end{array} \right.
\label{eqn:L2}
\end{equation}
\begin{equation}
\begin{array}{rcl}
\beta_n & = &
\displaystyle\frac{1}{(q^{\kappa -\iota +2 })_{2m}} \sum_{
           {\nu _1 \geq \cdots \geq \nu _{\kappa -1}
            \geq \nu _{\kappa }=0}
      \atop{\nu _1 + \cdots + \nu _{\kappa -1} \leq n}}
q^{\nu _1^2 + \cdots + \nu _{\kappa -1}^2
  +\nu _{\iota } + \cdots + \nu _{\kappa -1}} \\
& \times &
\displaystyle\prod_{\mu =1}^{\kappa -1}
\left[ \begin{array}{c}
2n+\kappa -\iota +1
-2(\nu _1 + \cdots + \nu _{\mu -1})
-\nu _{\mu }-\nu _{\mu +1} -\alpha^{(\kappa )}_{\iota \mu } \\
\nu _{\mu }-\nu _{\mu +1}
\end{array} \right]_q .
\end{array}
\label{eqn:R2}
\end{equation}
Applying Corollary \ref{cor:And} to
this Bailey pair we obtain the following fermionic expression
for the Virasoro character:
\begin{equation}
\begin{array}{rcl}
\chi^{(2\kappa +1, (2\kappa +1)k+2)}_{\iota ,k(\kappa +1)+1}(q)
& = &
\displaystyle\sum_{n_1 \geq \cdots \geq n_{k}\geq 0}
\frac{q^{n_1^2 \cdots + n_{k}^2 +
(\kappa -\iota +1)(n_1 + \cdots + n_k )}}
 {(q)_{n_1 - n_2 }\cdots (q)_{n_{k-1}-n_{k}}
  (q)_{2n_{k}+\kappa -\iota +1}} \\
& \times &
\displaystyle\sum_{
\nu_1 \geq \cdots \geq \nu_{\kappa -1} \geq \nu_{\kappa }=0 \atop
      \nu_{1} + \cdots + \nu_{\kappa -1} \leq n_{k}}
q^{\nu_1^2 + \cdots + \nu_{\kappa -1}^2
+\nu_{\iota } + \cdots + \nu_{\kappa -1}} \\
\times &
\displaystyle
\prod_{\mu =1}^{\kappa -1} &
\displaystyle\left[ \begin{array}{c}
2n_{k}+\kappa -\iota +1
-2(\nu_{1} + \cdots + \nu_{\mu -1})-\nu_{\mu } -\nu_{\mu +1}
-\alpha^{(\kappa )}_{\iota \mu } \\
\nu_{\mu } -\nu_{\mu +1}
\end{array} \right]_q .
\end{array}
\end{equation}
Furthermore, applying Corollary \ref{cor:AAB} to this pair
(\ref{eqn:L2}--\ref{eqn:R2}) we get
\begin{equation}
\begin{array}{cl}
& \displaystyle\sum_{j=0}^i
q^{j(\kappa -\iota +1)}
\chi^{(2\kappa +1,(2\kappa +1)k+2)}_{
       \iota ,(\kappa +1)k+1-i+2j}(q) \\
= &
\displaystyle\sum_{n_1 \geq \cdots \geq n_{k}\geq 0}
\frac{q^{n_1^2 \cdots + n_{k}^2 +
(\kappa -\iota )(n_1 +\cdots + n_i )+
(\kappa -\iota +1)(n_{i+1} + \cdots + n_k )}}
{(q)_{n_1 - n_2 }\cdots (q)_{n_{k-1}-n_{k}}
 (q)_{2n_{k}+\kappa -\iota +1}} \\
\times &
\displaystyle
\sum_{\nu _1 \geq \cdots \geq \nu _{\kappa -1}
      \geq \nu _{\kappa }=0 \atop
      \nu _{1} + \cdots + \nu _{\kappa -1} \leq n_{k}}
q^{\nu_1^2 + \cdots + \nu_{\kappa -1}^2
   +\nu _{\iota } + \cdots + \nu _{\kappa -1}} \\
\times &
\displaystyle
\prod_{\mu =1}^{\kappa -1}
\left[ \begin{array}{c}
2n_{k}+\kappa -\iota +1
-2(\nu _{1} + \cdots + \nu _{\mu -1})-\nu _{\mu }
-\nu _{\mu +1}
-\alpha^{(\kappa )}_{\iota \mu } \\
\nu _{\mu } -\nu _{\mu +1}
\end{array} \right]_q .
\end{array}
\label{eqn:ext}
\end{equation}
where $0 \leq i \leq k$. Notice that the LHS of the above equation
is a weighted sum of Virasoro characters. Extracting single character
from such an equation is beyond the scope of this work.

\subsection{Fermionic sum representations for
${\cal M}(2\kappa +1, (2\kappa +1)k+2\kappa -1)$}

The dual Bailey pair to (\ref{eqn:L1}--\ref{eqn:R1})
is
\begin{equation}
A_n = \left\{
\begin{array}{ll}
1, & n=0, \\
q^{r((2\kappa +1)(2\kappa -1)r+
   (2\kappa +1)(\kappa -\iota -1)+2\iota )},
& n=(2\kappa +1)r ~(r \geq 1), \\
q^{r((2\kappa +1)(2\kappa -1)r-
   (2\kappa +1)(\kappa -\iota -1)+2\iota )},
& n=(2\kappa +1)r-\kappa +\iota , ~(r \geq 1)\\
-q^{((2\kappa +1)r -\iota )
    ((2\kappa -1)r -\kappa +1)},
& n=(2\kappa +1)r-\kappa ~(r \geq 1), \\
-q^{((2\kappa +1)r +\iota )
    ((2\kappa -1)r +\kappa -1)},
& n=(2\kappa +1)r+\iota ~(r \geq 0), \\
0, & \mbox{otherwise},
\end{array} \right.
\label{eqn:L1D}
\end{equation}
\begin{equation}
\begin{array}{rcl}
B_n & = &
\displaystyle\frac{q^{n^2 +(\kappa -\iota )n}}
                  {(q^{\kappa -\iota +1})_{2n}} \sum_{
           {\nu_1 \geq \cdots \geq \nu_{l-1} \geq \nu_{\kappa }=0}
      \atop{\nu_1 + \cdots + \nu_{\kappa -1} \leq n}}
q^{\nu _1^2 + \cdots + \nu _{\kappa -1}^2 -
   \nu _1 (2m+\kappa -\iota )} \times \\
& \times &
\displaystyle\prod_{\mu =1}^{\kappa -1}
\left[ \begin{array}{c}
2n+\kappa -\iota -
2(\nu _1 + \cdots + \nu _{\mu -1})-\nu _{\mu }-\nu _{\mu +1}
-\alpha^{(\kappa )}_{\iota \mu } \\
\nu _{\mu }-\nu _{\mu +1}
\end{array} \right]_q .
\end{array}
\label{eqn:R1D}
\end{equation}
In order to obtain $B_m$, we use
$$
\left[ \begin{array}{c}
N+M \\M
\end{array} \right]_{q^{-1}}
=  q^{-NM}
\left[ \begin{array}{c}
N+M \\M
\end{array} \right]_{q},
$$
and
$$
\sum_{\mu =1}^{\kappa -1}
(\nu _{\mu }-\nu _{\mu +1})
(2(\nu _1 + \cdots + \nu _{\mu }) +\alpha^{(\kappa )}_{\iota \mu })
= 2 \sum_{\mu =1}^{\kappa -1} \nu _{\mu }^2
  + \sum_{\mu =\iota }^{\kappa -1} \nu _{\mu }.
$$
{}From
\begin{equation}
\frac{1}{(q)_{\infty}} \sum_{n=0}^{\infty }
A_n (q^{\kappa -\iota })^{kn} q^{kn^2 }
= \chi^{(2\kappa +1, (2\kappa +1)k + 2\kappa -1)}_{
        \iota , \kappa k +\kappa -1}(q),
\end{equation}
and Corollary \ref{cor:And},
we obtain
\begin{equation}
\begin{array}{rcl}
\chi^{(2\kappa +1, (2\kappa +1)(k+1)-2)}_{
      \iota, \kappa k +\kappa -1}(q)
& = &
\displaystyle\sum_{n_1 \geq \cdots \geq n_{k}\geq 0}
\frac{q^{n_1^2 +\cdots + n_{k-1}^2 +2n_{k}^2
        +(\kappa -\iota )(n_1 + \cdots + n_{k-1} +2n_{k})}}
     {(q)_{n_1 - n_2 }\cdots (q)_{n_{k-1}-n_{k}}
      (q)_{2n_{k}+\kappa -\iota }} \\
& \times &
\displaystyle\sum_{
\nu _1 \geq \cdots \geq \nu _{\kappa -1} \geq \nu _{\kappa }=0 \atop
\nu _{1} + \cdots + \nu _{\kappa -1} \leq n_{k}}
q^{\nu _1^2 + \cdots + \nu _{\kappa -1}^2 -
   \nu _1 (2n_k + \kappa -\iota )} \times \\
\times & \displaystyle\prod_{\mu =1}^{\kappa -1} &
\displaystyle\left[ \begin{array}{c}
2n_{k} +\kappa -\iota -2(\nu_{1} +
\cdots + \nu _{\mu -1})-\nu _{\mu } -\nu _{\mu +1}
-\alpha^{(\kappa )}_{\iota \mu } \\
\nu _{\mu } -\nu _{\mu +1}
\end{array} \right]_q .
\end{array}
\end{equation}

The dual Bailey pair to (\ref{eqn:L2}--\ref{eqn:R2}) is
\begin{equation}
A_n = \left\{
\begin{array}{ll}
1, & n=0, \\
q^{r((2\kappa +1)(2\kappa -1)r+
      (2\kappa +1)(\kappa -\iota )+2\iota )},
& n=(2\kappa +1)r ~(r \geq 1), \\
q^{r((2\kappa +1)(2\kappa -1)r-
      (2\kappa +1)(\kappa -\iota )+2\iota )},
& n=(2\kappa +1)r-\kappa +\iota -1 ~(r \geq 1), \\
-q^{((2\kappa -1)r +\kappa )((2\kappa +1)r +\iota )},
& n=(2\kappa +1)r +\iota ~(r \geq 0), \\
-q^{((2\kappa -1)r -\kappa )((2\kappa +1)r -\iota )},
& n=(2\kappa +1)r-\kappa -1 ~(r \geq 1), \\
0, & \mbox{otherwise},
\end{array} \right.
\label{eqn:L2p}
\end{equation}
\begin{equation}
\begin{array}{rcl}
B_n & = &
\displaystyle\frac{q^{n^2 +(\kappa -\iota +1)n}}
       {(q^{\kappa -\iota +2 })_{2n}} \sum_{
           {\nu _1 \geq \cdots \geq \nu _{\kappa -1}
            \geq \nu _{\kappa }=0}
      \atop{\nu _1 + \cdots + \nu _{\kappa -1} \leq n}}
q^{\nu _1^2 + \cdots + \nu _{\kappa -1}^2
  -\nu _{1} (2n+\kappa -\iota +1)} \\
& \times &
\displaystyle\prod_{\mu =1}^{\kappa -1}
\left[ \begin{array}{c}
2n+\kappa -\iota +1
-2(\nu _1 + \cdots + \nu _{\mu -1})
-\nu _{\mu }-\nu _{\mu +1} -\alpha^{(\kappa )}_{\iota \mu } \\
\nu _{\mu }-\nu _{\mu +1}
\end{array} \right]_q .
\end{array}
\label{eqn:R2p}
\end{equation}
Applying Corollary \ref{cor:And} to this Bailey pair we obtain the
following fermionic expression for the Virasoro character:
\begin{equation}
\begin{array}{rcl}
\chi^{(2\kappa +1, (2\kappa +1)k+2\kappa -1)}_{
       \iota ,k(\kappa +1)+\kappa }(q)
& = &
\displaystyle\sum_{n_1 \geq \cdots \geq n_{k}\geq 0}
\frac{q^{n_1^2 \cdots + n_{k-1}^2 + 2n_k^2 +
(\kappa -\iota +1)(n_1 + \cdots + n_{k-1} +2n_k )}}
 {(q)_{n_1 - n_2 }\cdots (q)_{n_{k-1}-n_{k}}
  (q)_{2n_{k}+\kappa -\iota +1}} \\
& \times &
\displaystyle\sum_{
\nu_1 \geq \cdots \geq \nu_{\kappa -1} \geq \nu_{\kappa }=0 \atop
      \nu_{1} + \cdots + \nu_{\kappa -1} \leq n_{k}}
q^{\nu_1^2 + \cdots + \nu_{\kappa -1}^2
-\nu_1 (\kappa -\iota +1)} \\
\times &
\displaystyle
\prod_{\mu =1}^{\kappa -1} &
\displaystyle\left[ \begin{array}{c}
2n_{k}+\kappa -\iota +1
-2(\nu_{1} + \cdots + \nu_{\mu -1})-\nu_{\mu } -\nu_{\mu +1}
-\alpha^{(\kappa )}_{\iota \mu } \\
\nu_{\mu } -\nu_{\mu +1}
\end{array} \right]_q .
\end{array}
\end{equation}
Furthermore, applying Corollary \ref{cor:AAB}
to this pair (\ref{eqn:L2p}--\ref{eqn:R2p}) we get
\begin{equation}
\begin{array}{cl}
& \displaystyle\sum_{j=0}^i
q^{j(\kappa -\iota +1)}
\chi^{(2\kappa +1,(2\kappa +1)k+2\kappa -1)}_{
       \iota ,(\kappa +1)k+\kappa -i+2j}(q) \\
= &
\displaystyle\sum_{n_1 \geq \cdots \geq n_{k}\geq 0}
\frac{q^{n_1^2 \cdots + n_{k-1}^2 + 2n_k^2 +
+(\kappa -\iota +1)(n_{1} + \cdots + n_{k-1} +2n_k )
-(n_1 +\cdots + n_i )}}
{(q)_{n_1 - n_2 }\cdots (q)_{n_{k-1}-n_{k}}
 (q)_{2n_{k}+\kappa -\iota +1}} \\
\times &
\displaystyle
\sum_{\nu _1 \geq \cdots \geq \nu _{\kappa -1}
      \geq \nu _{\kappa }=0 \atop
      \nu _{1} + \cdots + \nu _{\kappa -1} \leq n_{k}}
q^{n_1^2 + \cdots + n_{\kappa -1}^2
   +\nu _1 (2n_k +\kappa -\iota +1)} \\
\times &
\displaystyle
\prod_{\mu =1}^{\kappa -1}
\left[ \begin{array}{c}
2n_{k}+\kappa -\iota +1
-2(\nu _{1} + \cdots + \nu _{\mu -1})-\nu _{\mu }
-\nu _{\mu +1}
-\alpha^{(\kappa )}_{\iota \mu } \\
\nu _{\mu } -\nu _{\mu +1}
\end{array} \right]_q .
\end{array}
\end{equation}
where $0 \leq i \leq k$.

\section{Fermionic sum representations of the type (III)}

In this section we apply the Andrews--Bailey construction to the
polynomial identity obtained in \cite{Mel1,Ber}.

\subsection{Another polynomial identity}

Let $I_n $ and $C_n =2-I_n $ stand for the incidence
and Cartan matrices of the Lie algebra $A_n$, respectively:
$$
(I_n)_{ab} = \delta_{a,b+1}+\delta_{a,b-1}
{}~~~ a,b = 1, \cdots , n.
$$
Let ${\bf e}_a $ be the $a$th unit vector in $\bc ^n $
and set ${\bf e}_a = {\bf 0}$
for $a<0$ or $a>n$.
Define the following symbol
\begin{equation}
F^{(L)}_n \left[ \begin{array}{c}
{\bf Q} \\ {\bf A} \end{array} \right] ({\bf u} | q) =
\sum_{{\bf m}\in 2 \bz^n +{\bf Q}}
q^{({\bf m}, C_n {\bf m})/4 -({\bf A}, {\bf m})/2}
\prod_{a=1}^{n}
\left[
\begin{array}{c}
({\bf e}_a , I_n {\bf m} + {\bf u} +L{\bf e}_1)/2 \\
({\bf e}_a , {\bf m})
\end{array}
\right]_q ,
\end{equation}
where
$({\bf x},{\bf y})$ is the standard inner product
in $\bc ^n $, and
$$
\left[
\begin{array}{c}
N \\
M
\end{array}
\right]_q =
\left\{ \begin{array}{ll}
\displaystyle\frac{(q)_{N}}{(q)_{M}(q)_{N-M}}, &
\mbox{if $0 \leq M \leq N$, } \\
0, & {\rm otherwise. }
\end{array} \right.
$$

For $n=p-2 \geq 1$, set
$$
{\bf Q}_{r,s}=(s-1)({\bf e}_1 + \cdots + {\bf e}_n )
+({\bf e}_{r-1} +{\bf e}_{r-3} + \cdots )
+({\bf e}_{p+1-s} +{\bf e}_{p+3-s} + \cdots ).
$$
and define the following two $q$-series:
\begin{equation}
F^{(L)}_{p;r}(q) =
q^{-r(r-1)} \times
\left\{ \begin{array}{ll}
F^{(L)}_{p-2} \left[ \begin{array}{c}
{\bf Q}_{r,1} \\ {\bf 0} \end{array} \right]
({\bf e}_r | q) &
\mbox {if $L \not\equiv r-1$ mod $2$}, \\
F^{(L)}_{p-2} \left[ \begin{array}{c}
{\bf Q}_{p-r,p} \\ {\bf 0} \end{array} \right]
({\bf e}_{p-r} | q) &
\mbox{if $L \equiv r-1$ mod $2$},
\end{array}
\right.
\end{equation}
\begin{equation}
\overline{F}^{(L)}_{p;r}(q) =
q^{-r(r-1)} \times
\left\{ \begin{array}{ll}
F^{(L)}_{p-2} \left[ \begin{array}{c}
{\bf Q}_{r,1} \\ {\bf e}_r +L{\bf e}_1 \end{array} \right]
({\bf e}_r | q) &
\mbox{if $L \not\equiv r-1$ mod $2$}, \\
F^{(L)}_{p-2} \left[ \begin{array}{c}
{\bf Q}_{p-r,p} \\ {\bf e}_{p-r} +L{\bf e}_1  \end{array} \right]
({\bf e}_{p-r} | q) &
\mbox{if $L \equiv r-1$ mod $2$}.
\end{array}
\right.
\end{equation}

We also introduce
\begin{equation}
\begin{array}{rcl}
B^{(L)}_{p;r}(q) &=&
\displaystyle\sum_{j\in \bz } \left(
q^{j(jp(p+1)+r(p+1)-p)}
\left[ \begin{array}{c}
L \\
\left[ \frac{L-r+1}{2} \right] -j(p+1)
\end{array} \right]_q \right. \\
&-& \left.
q^{(jp+r)(j(p+1)+1)}
\left[ \begin{array}{c}
L \\
\left[ \frac{L-r-1}{2} \right] -j(p+1)
\end{array} \right]_q  \right).
\end{array}
\end{equation}

\begin{prop} The following polynomial identity holds:
\begin{equation}
B^{(L)}_{p;r}(q)=F^{(L)}_{p;r}(q).
\label{eqn:MB}
\end{equation}
\end{prop}

As for the proof see \cite{Ber}.

\subsection{Fermionic sum representations for
${\cal M}(p, kp+p-1)$}

In what follows we replace $p$ by $p-1$.
Accordingly, $p \geq 4$ and $1 \leq r \leq p-2$.
Set $L =2l+r-1 $ and divide (\ref{eqn:MB})
by $(q^{r})_{2l}$. Then
we can read the following Bailey pair relative to
$q^{r-1}$:
\begin{equation}
\alpha_l = \left\{
\begin{array}{ll}
1, & l=0, \\
q^{j(jp(p-1)+rp-p+1)}
& l=jp ~(j \geq 1), \\
q^{j(jp(p-1)-rp+p-1)}
& l=jp-r+1 ~(j \geq 1), \\
-q^{(j(p-1)+r)(jp+1)}
& l=jp +1 ~(j \geq 0), \\
-q^{(j(p-1)-r)(jp-1)}
& l=jp-r ~(j\geq 1), \\
0, & \mbox{otherwise},
\end{array} \right.
\label{eqn:L3}
\end{equation}
\begin{equation}
\begin{array}{rcl}
\beta_l & = &
 \displaystyle\frac{1}{(q^{r})_{2l}} F^{(2l+r-1)}_{p-1;r}(q).
\end{array}
\label{eqn:R3}
\end{equation}
Here and hereafter terms belonging to the same
congruence class modulo $p$ should be
summed up if exists. For instance, when $r=1$
(\ref{eqn:L3}) should read as follows:
\begin{equation}
\alpha_l = \left\{
\begin{array}{ll}
1, & l=0, \\
q^{j(jp(p-1)+rp-p+1)}
+q^{j(jp(p-1)-rp+p-1)}
& l=jp ~(j \geq 1), \\
-q^{(j(p-1)+r)(jp+1)}
& l=jp +1 ~(j \geq 0), \\
-q^{(j(p-1)-r)(jp-1)}
& l=jp-1 ~(j\geq 1), \\
0, & \mbox{otherwise},
\end{array} \right.
\end{equation}

By applying Lemma \ref{cor:And} to the Bailey pair
(\ref{eqn:L3}--\ref{eqn:R3}), we obtain the following
fermionic expression for the Virasoro character:
\begin{equation}
\begin{array}{rcl}
\chi^{(p, kp+p-1)}_{1,r(k+1)}(q)
& = &
\displaystyle\sum_{n_1 \geq \cdots \geq n_{k}\geq 0}
\frac{q^{n_1^2 +\cdots + n_{k}^2
        +(r-1)(n_1 + \cdots + n_{k})}}
     {(q)_{n_1 - n_2 }\cdots (q)_{n_{k-1}-n_{k}}
      (q)_{2n_{k}+r-1}} F^{(2n_k +r-1)}_{p-1;r}(q).
\end{array}
\label{eqn:k,kp+p-1}
\end{equation}

Substituting $L=2l+r$ into (\ref{eqn:MB}) and dividing by
$(q^{r+1})_{2l}$ we obtain another Bailey pair relative to
$q^{r}$
\begin{equation}
\alpha_l = \left\{
\begin{array}{ll}
1, & l=0, \\
q^{j(jp(p-1)+rp-p+1)}
& l=jp ~(j \geq 1), \\
q^{j(jp(p-1)-rp+p-1)}
& l=jp-r ~(j \geq 1), \\
-q^{(j(p-1)+r)(jp+1)}
& l=jp +1 ~(j \geq 0), \\
-q^{(j(p-1)-r)(jp-1)}
& l=jp-r-1 ~(j \geq 1), \\
0, & \mbox{otherwise},
\end{array} \right.
\label{eqn:L4}
\end{equation}
\begin{equation}
\begin{array}{rcl}
\beta_l & = &
 \displaystyle\frac{1}{(q^{r+1})_{2l}} F^{(2l+r)}_{p-1;r}(q).
\end{array}
\label{eqn:R4}
\end{equation}
Applying Lemma \ref{cor:And} to this Bailey pair we obtain the
following fermionic expression for the Virasoro character:
\begin{equation}
\begin{array}{rcl}
\chi^{(p, kp+p-1)}_{1,r(k+1)+k}(q)
& = &
\displaystyle\sum_{n_1 \geq \cdots \geq n_{k}\geq 0}
\frac{q^{n_1^2 +\cdots + n_{k}^2
        +r(n_1 + \cdots + n_{k})}}
     {(q)_{n_1 - n_2 }\cdots (q)_{n_{k-1}-n_{k}}
      (q)_{2n_{k}+r}} F^{(2n_k +r)}_{p-1;r}(q).
\end{array}
\label{eqn:k,kp+p-1,2}
\end{equation}
Furthermore, applying Corollary \ref{cor:AAB}
to this pair (\ref{eqn:L4}--\ref{eqn:R4}) we get
\begin{equation}
\sum_{j=0}^{i} q^{jr}
\chi^{(p, kp+p-1)}_{1,r(k+1)+k-i+2j}(q)
=
\displaystyle\sum_{n_1 \geq \cdots \geq n_{k}\geq 0}
\frac{q^{n_1^2 +\cdots + n_{k}^2
        +(r-1)(n_1 + \cdots + n_{i})
        +r(n_{i+1} + \cdots + n_{k})}}
     {(q)_{n_1 - n_2 }\cdots (q)_{n_{k-1}-n_{k}}
      (q)_{2n_{k}+r}} F^{(2n_k +r)}_{p-1;r}(q).
\end{equation}

\subsection{Fermionic sum representations for ${\cal M}(k,kp+1)$}

The dual Bailey pair to (\ref{eqn:L3}--\ref{eqn:R3})
is
\begin{equation}
A_l = \left\{
\begin{array}{ll}
1, & n=0, \\
q^{j(jp-1)}
& l=jp ~(j \geq 1), \\
q^{j(jp+1)}
& l=jp-r+1 ~(j \geq 1), \\
-q^{j(jp+1)}
& l=jp+1 ~(j \geq 0), \\
-q^{j(jp-1)}
& l=jp-r ~(j \geq 1), \\
0 & \mbox{otherwise},
\end{array} \right.
\label{eqn:L3D}
\end{equation}
\begin{equation}
B_l = \frac{q^{l^2 +(r-1)l}}{(q^r )_{2l}}
\overline{F}^{(2l+r-1)}_{p-1;r}(q).
\label{eqn:R3D}
\end{equation}
{}From
\begin{equation}
\frac{1}{(q)_{\infty}} \sum_{l=0}^{\infty }
A_l q^{(r-1)kl} q^{kl^2 }
= \chi^{(p,kp+1)}_{1,kr}(q),
\end{equation}
and Lemma \ref{cor:And},
we obtain
\begin{equation}
\chi^{(p,kp+1)}_{1,kr}(q)
= \sum_{n_1 \geq \cdots \geq n_{k}\geq 0}
\frac{q^{n_1^2 +\cdots + n_{k-1}^2 +2n_k^2
        +(r-1)(n_1 + \cdots + n_{k-1}+2n_k )}}
     {(q)_{n_1 - n_2 }\cdots (q)_{n_{k-1}-n_{k}}
      (q)_{2n_{k}+r-1}}
\overline{F}^{(2n_k +r-1)}_{p-1;r}(q) .
\label{eqn:SB}
\end{equation}
Setting $r=1$ and
$n_i -n_{i+1} =\nu_i $ ($i=1, \cdots , k-1$),
$2n_k =\nu_k$, $m_a =\nu_{k+a}$ ($a=1, \cdots , p-3$)
in (\ref{eqn:SB}),
we reproduce the corresponding expressions given in \cite{SB1}.

The dual Bailey pair to (\ref{eqn:L4}--\ref{eqn:R4})
is
\begin{equation}
A_l = \left\{
\begin{array}{ll}
1, & n=0, \\
q^{j(jp+p-1)}
& l=jp ~(j \geq 1), \\
q^{j(jp-p+1)}
& l=jp-r ~(j \geq 1), \\
-q^{(j+1)(jp+1)}
& l=jp+1 ~(j \geq 0), \\
-q^{(j-1)(jp-1)}
& l=jp-r ~(j \geq 1), \\
0, & \mbox{otherwise},
\end{array} \right.
\label{eqn:L4d}
\end{equation}
\begin{equation}
\begin{array}{rcl}
B_l & = &
\displaystyle\frac{q^{l^2 +rl}}{(q^{r+1})_{2l}}
\overline{F}^{(2l+r)}_{p-1;r}(q).
\end{array}
\label{eqn:R4d}
\end{equation}
Applying Lemma \ref{cor:And} to this Bailey pair we obtain the
following fermionic expression for the Virasoro character:
\begin{equation}
\begin{array}{rcl}
\chi^{(p,kp+1)}_{1,k(r+1)+1}(q)
& = &
\displaystyle\sum_{n_1 \geq \cdots \geq n_{k}\geq 0}
\frac{q^{n_1^2 +\cdots + n_{k-1}^2 + 2n_k^2
        +r(n_1 + \cdots + n_{k-1}+2n_k )}}
     {(q)_{n_1 - n_2 }\cdots (q)_{n_{k-1}-n_{k}}
      (q)_{2n_{k}+r}}
\overline{F}^{(2n_k +r)}_{p-1;r}(q).
\end{array}
\end{equation}
Furthermore, applying Corollary \ref{cor:AAB} to this pair
(\ref{eqn:L4d}--\ref{eqn:R4d}) we get
\begin{equation}
\sum_{j=0}^{i} q^{jr}
\chi^{(p,kp+1)}_{1,k(r+1)+1-i+2j}(q)
=
\displaystyle\sum_{n_1 \geq \cdots \geq n_{k}\geq 0}
\frac{q^{n_1^2 +\cdots + n_{k-1}^2 + 2n_k^2
        +r(n_1 + \cdots + n_{k-1}+2n_k )
        -(n_1 + \cdots + n_i)}}
     {(q)_{n_1 - n_2 }\cdots (q)_{n_{k-1}-n_{k}}
      (q)_{2n_{k}+r}}
\overline{F}^{(2n_k +r)}_{p-1;r}(q).
\end{equation}

\section{Concluding remarks}

In section 3 we derived the fermionic sum representations for
Virasoro characters listed in (II) in section 1. In section 4,
we derived those listed in (III). The point of this paper is
to point out that the Andrews--Bailey construction provides
a systematic way to prove infinite families of Viarasoro
character identities. For instance, in order to derive $q$-series
identitiies related to (II) and (III), we applied the Andrews--Bailey
construction to polynomial identities obtained in \cite{FQ} and
\cite{Mel1,Ber}, respectively. The starting point is always a suitable
Bailey pair, or equivalently a polynomial identity. Once the latter
is established, the proof of an infinite family of Virasoro character
identities becomes a matter of straightforward computation. In
\cite[Theorem 1.2]{FQ} we proved another polynomial identity.
Unfortunately, however, the application of the Andrews--Bailey
construction to this polynomial identity dose not provide us any
fruitful results.

We would like to mention fermionic representations of the type (III).
In the last section we restricted ourselves $p \geq 4$.
When $p=2$, the equivalence between bosonic and fermionic
representations are nothing but Gordon's generalization of
the Rogers--Ramanujan identities \cite{Andbk}. Interpretation
of Gordon's identity in terms of that equivalence was presented in
\cite{FNO,NRT}. The case $p=3$ of (III) is rather realized by
setting $\iota =\kappa =1$ in (II).

Proving (i) for any $p$, $r$ and $s$ is a still open problem
\footnote{ Berkovich has recently succeeded in proving (i) for
any $p$, $r$ and $s$ \cite{Ber-pri}. }. When this can be solved,
the list (III) will be extended to

\begin{tabular}{cl}
($\widehat{\rm III}$) & $\left\{
\begin{tabular}{l}
$\chi^{(p,kp+p-1)}_{s,r(k+1)}(q)$ and
$\chi^{(p,kp+p-1)}_{s,r(k+1)+k}(q)$ \\
$\chi^{(p,kp+1)}_{s,kr}(q)$ and
$\chi^{(p,kp+1)}_{s,k(r+1)+1}(q)$
\end{tabular} \right.$
for $p \geq 4$, $1 \leq r \leq p-2,
1 \leq s \leq p-1$ and $k\geq 1$.
\end{tabular}

\noindent Furthermore, there exist a number of curious phenomena
reported in \cite{SB2,KNS,KR,KRV,Mel2}. We wish to discuss these
matters in a separate paper.

\section*{Acknowledgment}
We would like to thank Professors G. E. Andrews, A. Berkovich,
B. M. McCoy, E. Melzer, P. A. Pearce,  and M. R\"{o}sgen, S. O.
Warnaar, and Y.-K. Zhou for useful information and interest in
this work. This work is supported by the Australian Research
Council.

\appendix

\section{Virasoro characters and Bailey pairs}

In this appendix, we list
several fermionic representations
obtained from Slater's table \cite{Sla}.
Here even when $p$ and $p'$ are not coprime,
we denote the infinite sum defined by the RHS of (\ref{eqn:RC})
by $\chi^{(p,p')}_{r,s}(q)$.

Let us begin by considering
the B and E series in Slater's table \cite{Sla}.
In each case $\alpha_0 =1$.

{}~

\begin{tabular}{|c|c|l|c|} \hline
 & $a$ & $\alpha_n $ & $ \beta_n $ \\ \hline
B(1) & $1$ &
$(-1)^n q^{3n^2 /2}(q^{n/2}+q^{-n/2})$ &
$1/(q)_n$ \\
B(2) & $1$ &
$(-1)^n q^{3n^2 /2}(q^{3n/2}+q^{-3n/2})$ &
$q^n /(q)_n $ \\
B(3) & $q$ &
$(-1)^n q^{(3n^2 +n)/2}(1-q^{2n+1})/(1-q)$ &
$1/(q)_n$ \\
E(1) & $1$ &
$2(-1)^n q^{n^2}$ &
$1/((q)_n (-q)_n )$ \\
E(3) & $q$ &
$(-1)^n q^{n^2 }(1-q^{2n+1})/(1-q)$ &
$1/((q)_n (-q)_n)$ \\
E(4) & $1$ &
$(-1)^n q^{n^2}(q^{n}+q^{-n})$ &
$q^n /((q)_n (-q)_n )$ \\
\hline
\end{tabular}

{}~

By applying Corollary \ref{cor:And} to these Bailey pairs,
we have
\begin{equation}
\begin{array}{cl}
{\rm B(1)} &
\displaystyle\chi^{(2,2k+3)}_{1,k+1}(q)=
\sum_{n_1 \geq \cdots \geq n_k \geq 0}
\frac{q^{n_1^2 +\cdots + n_{k}^2 }}
     {(q)_{n_1 -n_2 }\cdots (q)_{n_{k-1}-n_{k}} (q)_{n_k}}. \\
{\rm B(2)} &
\displaystyle\chi^{(2,2k+3)}_{1,k}(q)=
\sum_{n_1 \geq \cdots \geq n_k \geq 0}
\frac{q^{n_1^2 +\cdots + n_{k}^2 +n_k }}
     {(q)_{n_1 -n_2 }\cdots (q)_{n_{k-1}-n_{k}} (q)_{n_k}}. \\
{\rm B(3)} &
\displaystyle\chi^{(2,2k+3)}_{1,1}(q)=
\sum_{n_1 \geq \cdots \geq n_k \geq 0}
\frac{q^{n_1^2 +\cdots + n_{k}^2 + n_1 +\cdots + n_k }}
     {(q)_{n_1 -n_2 }\cdots (q)_{n_{k-1}-n_{k}} (q)_{n_k}}. \\
{\rm E(1)} &
\displaystyle\chi^{(2,2k+2)}_{1,k+1}(q)=
\sum_{n_1 \geq \cdots \geq n_k \geq 0}
\frac{q^{n_1^2 +\cdots + n_k^2 } }
     {(q)_{n_1 -n_2 }\cdots (q)_{n_{k-1}-n_{k}} (q)_{n_k }(-q)_{n_k }}. \\
{\rm E(3)} &
\displaystyle\chi^{(2,2k+2)}_{1,1}(q)=
\sum_{n_1 \geq \cdots \geq n_k \geq 0}
\frac{q^{n_1^2 +\cdots + n_{k}^2 + n_1 +\cdots + n_k }}
     {(q)_{n_1 -n_2 }\cdots (q)_{n_{k-1}-n_{k}} (q)_{n_k}(-q)_{n_k}}. \\
{\rm E(4)} &
\displaystyle\chi^{(2,2k+2)}_{1,k}(q)=
\displaystyle\chi^{(4,k+1)}_{1,k/2}(q)=
\sum_{n_1 \geq \cdots \geq n_k \geq 0}
\frac{q^{n_1^2 +\cdots + n_k^2 +n_k } }
     {(q)_{n_1 -n_2 }\cdots (q)_{n_{k-1}-n_{k}} (q)_{n_k }(-q)_{n_k }}.
\end{array}
\end{equation}

Furthermore, by applying Corollary \ref{cor:AAB}
to B(3), we obtain the interpolating expression
between B(1) and B(3):
\begin{equation}
\chi^{(2, 2k+3)}_{1,i}(q) =
\chi^{(2, 2k+3)}_{1,2k+3-i}(q)
=
\displaystyle\sum_{n_1 \geq \cdots \geq n_{k} \geq 0}
\frac{q^{n_1^2 +\cdots + n_{k}^2 +n_i +\cdots +n_{k}}}
     {(q)_{n_1 -n_2 } \cdots (q)_{n_{k-1}-n_{k}} (q)_{n_{k}}},
\end{equation}
where $1 \leq i \leq k+1$.

{}From each form of $(\alpha , \beta )$, we can easily
guess that E(1), E(4) and E(3) are the counterpart of
B(1), B(2) and B(3). Actually,
using Corollary \ref{cor:AAB} we get a
interpolating expression between
E(1) and E(3)
\begin{equation}
\chi^{(2, 2k+2)}_{1, i+1}(q)
=
\chi^{(4, k+1)}_{2, (i+1)/2}(q)
=
\displaystyle\sum_{n_1 \geq \cdots \geq n_{k} \geq 0}
\frac{q^{n_1^2 +\cdots + n_{k}^2 +n_{i+1} +\cdots +n_{k}}}
     {(q)_{n_1 -n_2 } \cdots (q)_{n_{k-1}-n_{k}}
      (q)_{n_{k}}(-q)_{n_{k}}}.
\end{equation}
where $0 \leq i \leq k$.

Let us go on A series in her table \cite{Sla}.
In each case $\alpha_0 =1$.

{}~

\begin{tabular}{|c|c|l|l|l|c|} \hline
 & $a$ &
$\alpha_{3r}$ & $\alpha_{3r-1}$ & $\alpha_{3r+1}$
& $\beta_n $ \\ \hline
A(1) & $1$ &
$q^{6r^2 -r}+q^{6r^2 +r}$ &
$-q^{6r^2 -5r+1}$ &
$-q^{6r^2 +5r+1}$ &
$1/(q)_{2n}$ \\
A(2) & $q$ &
$q^{6r^2 +r}$ &
$~~q^{6r^2 -r}$ &
$-q^{6r^2 +5r+1} -q^{6r^2 +7r+2}$ &
$1/(q^2 )_{2n}$ \\
A(3) & $1$ &
$q^{6r^2 -2r}+q^{6r^2 +2r}$ &
$-q^{6r^2 -2r}$ &
$-q^{6r^2 +2r}$ &
$q^n /(q)_{2n}$ \\
A(4) & $q$ &
$q^{6r^2 +4r}$ &
$~~q^{6r^2 -4r}$ &
$-q^{6r^2 +4r} -q^{6r^2 +8r+2}$ &
$q^n /(q^2 )_{2n}$ \\
A(5) & $1$ &
$q^{3r^2 -r}+q^{3r^2 +r}$ &
$-q^{3r^2 -r}$ &
$-q^{3r^2 +r}$ &
$q^{n^2 }/(q)_{2n}$ \\
A(6) & $q$ &
$q^{3r^2 -r}$ &
$~~q^{3r^2 +r}$ &
$-q^{3r^2 +r} -q^{3r^2 +5r+2}$ &
$q^{n^2 }/(q^2 )_{2n}$ \\
A(7) & $1$ &
$q^{3r^2 -2r}+q^{3r^2 +2r}$ &
$-q^{3r^2 -4r+1}$ &
$-q^{6r^2 +4r+1}$ &
$q^{n^2 -n}/(q)_{2n}$ \\
A(8) & $q$ &
$q^{3r^2 +2r}$ &
$~~q^{3r^2 -2r}$ &
$-q^{2r^2 +2r} -q^{3r^2 +4r+1}$ &
$q^{n^2 +n}/(q^2 )_{2n}$ \\ \hline
\end{tabular}

{}~

Andrews \cite{And}
pointed out that Slater's A series consist of four
Bailey pairs and their dual pairs.
In fact, A(1) and A(5) are dual, the same holds between
A(2) and A(8); A(3) and A(7); A(4) and A(6).

By applying Corollary \ref{cor:And} to
A(1)--A(8) we obtain
\begin{equation}
\begin{array}{cl}
{\rm A(1)} &
\displaystyle\chi^{(3,3k+2)}_{1,k+1}(q)=
\sum_{n_1 \geq \cdots \geq n_k \geq 0}
\frac{q^{n_1^2 +\cdots + n_k^2 } }
     {(q)_{n_1 -n_2 }\cdots (q)_{n_{k-1}-n_{k}} (q)_{2n_k}}. \\
{\rm A(2)} &
\displaystyle\chi^{(3,3k+2)}_{1,2k+1}(q)=
\sum_{n_1 \geq \cdots \geq n_k \geq 0}
\frac{q^{n_1^2 +\cdots + n_k^2 +n_1 +\cdots + n_k } }
     {(q)_{n_1 -n_2 }\cdots (q)_{n_{k-1}-n_{k}} (q)_{2n_k +1}}. \\
{\rm A(3)} &
\displaystyle\chi^{(3,3k+2)}_{1,k}(q)=
\sum_{n_1 \geq \cdots \geq n_k \geq 0}
\frac{q^{n_1^2 +\cdots + n_k^2 +n_k } }
     {(q)_{n_1 -n_2 }\cdots (q)_{n_{k-1}-n_{k}} (q)_{2n_k }}. \\
{\rm A(4)} &
\displaystyle\chi^{(3,3k+2)}_{1,2k+2}(q)=
\sum_{n_1 \geq \cdots \geq n_k \geq 0}
\frac{q^{n_1^2 +\cdots + n_k^2 +n_1 +\cdots + n_{k-1}+ 2n_k } }
     {(q)_{n_1 -n_2 }\cdots (q)_{n_{k-1}-n_{k}} (q)_{2n_k +1}}. \\
{\rm A(5)} &
\displaystyle\chi^{(3,3k+1)}_{1,k}(q)=
\sum_{n_1 \geq \cdots \geq n_k \geq 0}
\frac{q^{n_1^2 +\cdots + n_{k-1}^2 +2n_k^2 } }
     {(q)_{n_1 -n_2 }\cdots (q)_{n_{k-1}-n_{k}} (q)_{2n_k }}. \\
{\rm A(6)} &
\displaystyle\chi^{(3,3k+1)}_{1,2k}(q)=
\sum_{n_1 \geq \cdots \geq n_k \geq 0}
\frac{q^{n_1^2 +\cdots + n_{k-1}^2 + 2n_k^2 +n_1 +\cdots + n_k } }
     {(q)_{n_1 -n_2 }\cdots (q)_{n_{k-1}-n_{k}} (q)_{2n_k +1}}. \\
{\rm A(7)} &
\displaystyle\chi^{(3,3k+1)}_{1,k+1}(q)=
\sum_{n_1 \geq \cdots \geq n_k \geq 0}
\frac{q^{n_1^2 +\cdots + n_{k-1}^2 +2n_k^2 -n_k } }
     {(q)_{n_1 -n_2 }\cdots (q)_{n_{k-1}-n_{k}} (q)_{2n_k }}. \\
{\rm A(8)} &
\displaystyle\chi^{(3,3k+1)}_{1,2k+1}(q)=
\sum_{n_1 \geq \cdots \geq n_k \geq 0}
\frac{q^{n_1^2 +\cdots + n_{k-1}^2 + 2n_k^2 +n_1 +\cdots + n_{k-1}+ 2n_k } }
     {(q)_{n_1 -n_2 }\cdots (q)_{n_{k-1}-n_{k}} (q)_{2n_k +1}}.
\end{array}
\end{equation}

Furthermore, by applying Corollary \ref{cor:AAB}
to A(2), we obtain the following expression
\begin{equation}
\sum_{j=0}^i
q^j \chi^{(3,3k+2)}_{1,2k-i+2j+1}(q)=
\sum_{n_1 \geq \cdots \geq n_k \geq 0}
\frac{q^{n_1^2 +\cdots + n_k^2 +n_{i+1} +\cdots + n_k } }
     {(q)_{n_1 -n_2 }\cdots (q)_{n_{k-1}-n_{k}} (q)_{2n_k +1}},
\label{eqn:3,3k+2}
\end{equation}
where $0 \leq i \leq k$.
We can reproduce these equation (\ref{eqn:3,3k+2})
by setting $\kappa =1$ in (\ref{eqn:ext}).


\begin{thebibliography}{99}
\bibitem{SB1}R. Kedem, T. R. Klassen, B. M. McCoy and E. Melzer,
Fermionic quasiparticle representations for characters of
$G^{(1)}_1 \times G^{(1)}_1 / G^{(1)}_2 $,
Phys. Lett. {\bf 304B} (1993) 263--270;
Fermionic sum representations for conformal field theory characters,
Phys. Lett. {\bf 307B} (1993) 68--76.
\bibitem{SB2}S. Dasmahapatra,
R. Kedem, T. R. Klassen, B. M. McCoy and E. Melzer,
Quasi-particles, conformal field theory, and
$q$-series, ITP-SB-93-12, hep-th/9303013,
in Proceedings of
{\it Yang--Baxter Equations in Paris},
J.-M. Maillard ed.;
R. Kedem, B. M. McCoy and E. Melzer,
The sums of Rogers, Schur and Ramanujan and
Bose--Fermi correspondence in $1+1$--dimensional
field theory, ITP-SB-93-19, hep-th/9304056.
\bibitem{ISZ}{\it Conformal Invariance and Application
to Statistical Mechanics}, C. Itzykson, H. Saleur
and J.-B. Zuber eds., World Scientific, 1988.
\bibitem{Mel1}E. Melzer,
Fermionic character sums and the
corner transfer matrix,
Int. J. Mod. Phys. {\bf A9} (1994) 1115--1136.
\bibitem{Ber}A. Berkovich, Fermionic counting of RSOS-states
and Virasoro character formulas for the unitary minimal
series ${\cal M}(\nu , \nu +1)$. Exact results,
hep-th/9403073, to appear in Nucl. Phys. B.
\bibitem{FQ}O. Foda and Y.-H. Quano,
Polynomial identities of the Rogers--Ramanujan type,
Univ. Melbourne preprint No.25, 1994, hep-th/9407191.
\bibitem{Andbk}G. E. Andrews, {\it The Theory of Partition},
Encyclopedia of Mathematics and its Application, Vol. 2,
G.-C. Rota ed., Addison-Wesley, 1976.
\bibitem{FNO}B. L. Feigin, T. Nakanishi and H. Ooguri,
The annihilating ideals of minimal models,
Int. J. Mod. Phys. {\bf A7} (Suppl. {\bf 1A}) (1992) 217--238.
\bibitem{NRT}W. Nahm, A. Recknagel and M. Terhoeven,
Dilogarithm identities in conformal field theory,
Mod. Phys. Lett. {\bf A8} (1993) 1835--1848.
\bibitem{Bai}W. N. Bailey, Some identities in combinatory analysis,
Proc. London Math. Soc. (2) {\bf 49} (1947) 421-435;
Identities of the Rogers--Ramanujan type,
Proc. London Math. Soc. (2) {\bf 50} (1949) 1-10.
\bibitem{And}G. E. Andrews, Multiple series
Rogers--Ramanujan type identities, Pac. J. Math. {\bf 114} (1984)
267--283.
\bibitem{AAB}A. K. Agarwal, G. E. Andrews and D. M. Bressoud,
The Bailey lattice, J. Indian Math. Soc. {\bf 51} (1987) 57--73.
\bibitem{Bre2}D. M. Bressoud, The Bailey lattice: An Introduction,
in {\it Ramanujan Revisited}, pp. 57-67,
G. E. Andrews et al. eds., Academic Press, Inc., 1988.
\bibitem{RC}A. Rocha-Caridi,
Vacuum vector representations of the Virasoro algebra,
in {\it Vertex Operators in Mathematics and Physics}, pp. 451--473,
J. Lepowsky et al. eds., Springer, 1985.
\bibitem{KNS}A. Kuniba, T. Nakanishi and J. Suzuki,
Characters in conformal field theories from thermodynamics Bethe ansatz,
Mod. Phys. Lett. {\bf A8} (1993) 1649--1660.
\bibitem{KR}J. Kellendonk and A. Recknagel,
Virasoro representations and fusion graphs,
Phys. Lett. {\bf 298B} (1993) 329--334.
\bibitem{KRV}J. Kellendonk, M. R\"{o}sgen and R. Varnhagen,
Path spaces and W fusion in minimal models,
Int. J. Mod. Phys. {\bf A9} (1994) 1009--1023.
\bibitem{Mel2}E. Melzer,
The many faces of a character,
Lett. Math. Phys. {\bf 31} (1994) 233--246.
\bibitem{GR}G. Gasper and M. Rahman,
{\it Basic Hypergeometric Series},
Encyclopedia of Mathematics and its Application, Vol. 35,
Addison-Wesley, 1990.
\bibitem{Sla}L. J. Slater, A new proof of Rogers's transformation of
infinite series, Proc. London Math. Soc. (2) {\bf 53}
(1951) 460--475;
Further identities of the Rogers--Ramanujan type,
Proc. London Math. Soc. (2) {\bf 54} (1952) 147--167.
\bibitem{LL}L. Lov\'{a}sz, {\it Combinatorial Problems and Exercises},
North Holland, 1979.
\bibitem{Ber-pri}A. Berkovich, private communication.
\end{thebibliography}
\end{document}